\documentclass[a4paper,fleqn,usenatbib]{mnras}

\usepackage{newtxtext,newtxmath}

\usepackage[T1]{fontenc}
\usepackage{ae,aecompl}

\usepackage{rotating}

\usepackage{graphicx}	
\usepackage{amsmath}	
\usepackage{amssymb}	
\usepackage{lscape}

\title[The FRIED grid of disc mass loss rates]{The FRIED grid of mass loss rates for externally irradiated protoplanetary discs}
\author[T. J. Haworth et al.]
{\parbox{\textwidth}{Thomas J. Haworth$^{1}$\thanks{E-mail: \texttt{t.haworth@imperial.ac.uk}}, Cathie J. Clarke$^2$, Wahidur Rahman$^1$, Andrew J. Winter$^2$ and Stefano Facchini$^3$  
}\vspace{0.4cm}\\
\parbox{\textwidth}{$^{1}$ Astrophysics Group, Imperial College London, Blackett Laboratory, Prince
Consort Road, London SW7 2AZ, UK \\
$^{2}$ Institute of Astronomy, Madingley Rd, Cambridge, CB3 0HA, UK \\
$^{3}$ Max-Planck-Institut f\"ur Extraterrestrische Physik, Giessenbachstrasse 1, 85748 Garching, Germany
}}

\begin{document}

\date{Accepted ???. Received ???; in original form ???}

\pagerange{\pageref{firstpage}--\pageref{lastpage}} \pubyear{2016}

\maketitle
\label{firstpage}

\begin{abstract}
We present an open access grid of 3930 calculations of externally evaporating protoplanetary discs. This spans a range of disc sizes (1--400\,AU), disc masses, UV field strengths (10--10$^4$\,G$_0$) and stellar masses  (0.05--1.9\,M$_\odot$). The grid is publicly available for download, and offers a means of cheaply including external photoevaporation in disc evolutionary calculations. It can also be queried using an online tool for quick estimates of instantaneous mass loss rates (e.g for convenient evaluation of real observed systems). \textsc{fried} itself illustrates that for discs around stars $\leq0.3$\,M$_\odot$ external photoevaporation is effective down to small radii ($<50$\,AU) down to UV fields at least as weak as 10\,G$_0$. At the other end of the scale, in a $10^4$\,G$_0$ environment photoevaporation is effective down to 1\,AU even for stellar masses at least as high as 1.9\,M$_\odot$. We also illustrate in which regimes CO survives in the photoevaporative outflow for significant mass loss rates; marking a system a good candidate to detect external photoevaporation in weak--intermediate UV environments through sub--Keplerian rotation. Finally we make illustrative mass loss rate estimates for discs in Taurus based on the \cite{2011A&A...529A.105G} star--disc parameters, finding that around half are expected to have both significant mass loss and retain CO in the photoevaporative outflow.

\end{abstract}


\section{introduction}
Planets are now known to exist around most stars (at least in the relatively local Milky Way) and exhibit a diverse range of architectures \citep{2015ARA&A..53..409W}. One of the key drivers in modern astrophysics is to understand the reason for this diversity, as well as how our own Solar system fits in to the wider population. To do so we must understand how the circumstellar ``protoplanetary'' discs of material around young stars not only ubiquitously give rise to planet formation, but also do so in a way that leads to high diversity in the resulting planetary parameters. To this end, substantial advances in our observational and theoretical capabilities have been made in recent years. Multi-wavelength observations from instruments like SPHERE \citep[e.g.][]{2017Msngr.169...32G}, ALMA \citep[e.g.][]{2015ApJ...808L...3A, 2017ApJ...851L..23C, 2018A&A...610A..24F} and the GPI \citep[e.g.][]{2041-8205-815-2-L26, 2041-8205-814-2-L27} are giving us the best observational insight yet into the inner workings of protoplanetary discs and planet formation \citep[for reviews and discussion on future advances see][]{2011ARA&A..49...67W, 2015PASP..127..961A, 2016PASA...33...59S}. Similarly, theoretical models are making rapid advances to capture the rich physics of planet-forming discs, which includes chemistry, magnetic fields, dust and gas dynamics and radiation transport \citep[for reviews and discussion of current and future advances in modelling of protoplanetary discs see][]{2016PASA...33...53H, 2016JGRE..121.1962M}. 

However, the problem is complicated further in that protoplanetary discs are found around young stars \citep[the discs are typically dispersed well before 10Myr, e.g.][]{2015A&A...576A..52R} and young stars are typically still in the clusters from which they formed. There are therefore \textit{environmental} factors that need to be accounted for.  These are: one-off gravitational encounters, a binary (or tertiary, etc.) companion,  and irradiation of a disc by other stellar members of the young cluster.

There is growing evidence that gravitational encounters are generally of secondary importance to photoevaporation. For example \cite{2001MNRAS.325..449S} demonstrated this when comparing dynamical and radiative distruption of discs in an Orion like environment.  Furthermore, \cite{2018MNRAS.475.2314W} and \cite{2018MNRAS.478.2700W} demonstrate that dynamical interactions are, statistically speaking, always a secondary effect in sculpting a disc population compared to photoevaporation, even in much weaker UV environments. Interactions have to be very close in order to see a significant effect on the disc evolution \citep{2018MNRAS.475.2314W}. Protoplanetary discs can more obviously be affected by a binary companion, potentially being significantly disrupted as in the case of RW Aurigae \citep[][]{2006A&A...452..897C, 2015MNRAS.449.1996D, 2018ApJ...859..150R} truncated \citep{1980ApJ...241..425G} or warped \citep{1983MNRAS.202.1181P, 2018MNRAS.473.4459F}. There is also evidence that this affects the resulting planetary populations \citep[e.g.][]{2002ApJ...568L.113Z, 2007A&A...462..345D}. 

The external photoevaporation of protoplanetary discs has been a challenging effect to gauge observationally.  For many years, only the proplyds -- discs within close proximity of O stars -- were obviously observed to be photoevaporating \citep[e.g.][]{1996AJ....111.1977M, 1998AJ....115..263O, 2001AJ....122.2662O, 2002ApJ...566..315H}. These are typically irradiated by a UV field of order $10^5$\,G$_0$\footnote{G$_0$ is the Habing unit of UV radiation, which is $1.6\times 10^{-3}$ erg cm$^{-2}$ s$^{-1}$ over the wavelength range ($912$\AA$<\lambda<2400$\AA)  \cite{1968BAN....19..421H}}. The effects of photoevaporation in the vicinity of O stars have also been studied recently by, for example, \cite{2017AJ....153..240A}, \cite{2018ApJ...860...77E} using the spatial distribution of disc properties. Some direct measures of the mass loss rate from discs near O stars have been made, for example by \cite{1987ApJ...321..516C}, \cite{1999AJ....118.2350H} and \cite{2002ApJ...566..315H}, finding mass loss rates of order $10^{-6}$\,M$_\odot$\,yr$^{-1}$.

In recent years there has been growing evidence of external photoevaporation in weaker radiation environments.  \cite{2016ApJ...826L..15K} found proplyds in a $\sim3\times10^3$\,G$_0$ environment and  \cite{2017MNRAS.468L.108H} used numerical models to propose external photoevaporation as the reason for the large CO halo around IM Lup, which is only in a $\sim4$\,G$_0$ environment \citep{2016ApJ...832..110C}. Nevertheless, external photoevaporation in weak--intermediate radiation environments is generally unconstrained, in part because it has not yet been actively searched for since the signatures of external photoevaporation aren't well known. As most stars are not in such a strong UV environment as the proplyds \citep{2008ApJ...675.1361F} understanding the evolution of discs in weaker environments is important.

Modelling the external photoevaporation of discs is difficult because the thermodynamic properties of the flow are set by photodissociation physics for the UV fields that the majority of star--discs are exposed to \citep[in the limit of being very close to a strong EUV source, photoionisation dominates, see Figure 12 of ][]{2018MNRAS.478.2700W}. Computing the thermal structure of a photodissociation region (PDR) requires the solution of a chemical network that is also sensitive to the non-local distribution of matter. That is, the temperature at one point in the flow is sensitive to the rest of the flow structure because this sets the cooling by the escape of line photons and also the
attenuation of the UV field by dust/molecular self-shielding
in outward lying regions of the photoevaporative flow. For this reason, for a long time only semi-analytic models of the flow structure and hence mass loss rate could be produced \citep[e.g.][]{1994ApJ...428..654H, 1998ApJ...499..758J, 2004ApJ...611..360A, 2016MNRAS.457.3593F}. These are quick to compute but are only able to obtain solutions in certain subsets of parameter space. To date they also all only consider 1\,M$_\odot$ stars. 

Computing solutions for arbitrary parts of the parameter space requires full photochemical-dynamical models that iteratively solve the PDR chemistry/temperature with the dynamics. This is both difficult to implement and computationally expensive, but has now been achieved by \cite{2016MNRAS.463.3616H} using the \textsc{torus-3dpdr} code (discussed in  section \ref{sec:num_meth}). Because these mass loss rates are difficult to compute, and expensive, having a large grid of publicly available pre-computed models would therefore open up consideration of external photoevaporation to the wider community. 

The value of pre-computed mass loss rates from models such as the above is that they can either be used to estimate the instantaneous mass loss rate for real systems, or can be applied to viscous evolutionary models of discs \citep[][]{2007MNRAS.376.1350C, 2013ApJ...774....9A, 2015ApJ...815..112K, 2017MNRAS.468L.108H, 2018MNRAS.475.5460H, 2018MNRAS.478.2700W}.  

In this paper we present the results of a large grid of external photoevaporation models as described above, which we refer to as the \textsc{fried} (\textbf{F}UV \textbf{R}adiation \textbf{I}nduced \textbf{E}vaporation of \textbf{D}iscs) grid. This covers a wide parameter space of stellar mass, disc mass, disc radius and UV field. It is publicly available for direct download, but we also provide an online tool for making quick mass loss rate estimates. The rest of this paper is as follows. In section \ref{sec:construction} we discuss how the \textsc{fried} grid is constructed, in section \ref{sec:theFriedGrid} we provide an overview of the resulting grid and the online resources. Finally in section \ref{sec:discussion} we apply the grid to an illustrative population of discs. 

\section{Preamble: easy access summary}
Our aim is for the grid of models in this paper to be widely used, and for both theoretical and observational applications. Since the models themselves are rather technical we therefore begin with a more accessible summary of the grid. 

We have calculated the mass loss rate of gas \citep[the mass loss rate of dust is related to, but different from the gas mass loss rate,][]{2016MNRAS.457.3593F, 2018MNRAS.475.5460H} from protoplanetary discs that are stripped of material due to external irradiation by nearby stars.  We have done this for a large variety of stellar/disc parameters as well as UV field strengths and tabulated this in the \textsc{fried} grid. Estimating this mass loss rate directly (without our grid) is both technically and computationally challenging, and so isn't usually considered either observationally or in numerical models. 

The grid itself is summarised in Figures \ref{fig:all10G0}-\ref{fig:all10000G0}, which each show the mass loss rate (colour scale) as a function of disc size ($y-$axis), disc mass ($x-$axis), and stellar mass in UV environments of $10$, $10^2$, $10^3$, $5\times10^3$ and $10^4$\,G$_0$ respectively. Each panel in these Figures represents a different stellar mass. Having these mass loss rate estimates over such a large parameter space allows estimates of the instantaneous mass loss rate to be made for real observed systems. That is, if you know the stellar mass, disc mass and disc radius you could estimate the mass loss rate as a function of the UV field (and similarly for other combinations of known/unknown parameters). 

To use the grid, the most efficient and flexible way is to download it in its entirety\footnote{\url{http://www.friedgrid.com/Downloads/}}. There is also an online tool for which you can provide the star--disc parameters and it will interpolate and return a mass loss rate\footnote{\url{http://www.friedgrid.com/Tool/}} (discussed in more detail in section \ref{sec:onlinetool}). 

At present \textsc{fried} includes only a single metallicity/PAH option, but can expand to accommodate this in future releases. 

\section{constructing the fried grid}
\label{sec:construction}
Here we describe the construction of our grid of photoevaporation models in detail. We begin by discussing the details of any given individual calculation and then discuss the coverage of the grid in section \ref{sec:params}.

\subsection{Photoevaporation calculations}
\label{sec:num_meth}
The disc photoevaporation calculations are computed using the \textsc{torus-3dpdr} extension \citep{2015MNRAS.454.2828B} of the \textsc{torus} Monte Carlo radiation transport and hydrodynamics code \citep{2000MNRAS.315..722H, 2015MNRAS.453.2277H, 2018MNRAS.477.5422A}. The approach is very similar to that first detailed and benchmarked in  \cite{2016MNRAS.463.3616H}.

The models involve iteratively performing hydrodynamics steps over some time interval $\Delta t$ and PDR chemistry calculations.  The thermal properties set by the PDR calculation then set the pressure distribution in the dynamical step. The calculations in this paper are 1D spherical, with the bulk of the mass loss assumed to be driven from the disc outer edge, where there is a large mass reservoir that is least gravitationally bound to the star \citep[see][for a discussion on this]{2004ApJ...611..360A}. Under such a scheme the mass loss rate is
 \begin{equation}
	\dot{M}_w = 4\pi R^2 \rho \dot{R} \mathcal{F}
	\label{equn:mdot}
\end{equation}
where  $\dot{R}$  is the flow velocity, with density $\rho$ at distance $R$, and $\mathcal{F}$ is the fraction of solid angle subtended by the disc outer edge $R_d$, which \cite{2004ApJ...611..360A} define as
\begin{equation}
	\mathcal{F} = \frac{H_d}{\sqrt{H_d^2+R_{d}^2}}
\end{equation}
The hydrodynamics itself is a grid based finite volume scheme with a point source gravitational potential set by the parent star. We generally use a \cite{vanleer} flux limiter and a Courant--Friedrichs--Lewy parameter of 0.3. 

The PDR calculation considers 33 species and 330 reactions, and was tailored to give temperatures accurate to within around 10\,per cent of the UMIST 2012 chemical network database of 215 species and over 3000 reactions \citep{2013A&A...550A..36M}. A summary of the species included and the initial abundances assumed is given in Table \ref{table:speciesparams}. {Note that chemical equilibrium is assumed, so the initial abundances are unimportant other than to set the initial distribution of metals and computational time taken to reach equilibrium. In the models of \cite{2016MNRAS.463.3616H}, which are similar to those here, we justified the assumption of equilibrium by showing that the thermal timescale was indeed faster than the flow timescale. Recently, time-dependent models tailored to the study the FUV internal photoevaporation of discs by the host star have been computed by \cite{2017ApJ...847...11W} and \cite{2018ApJ...857...57N}. } 

{The main cooling in these models is the escape of line photons of C\,I, C\,II, O\,I and CO \citep[see the right hand panel in Figure 2 of][]{2016MNRAS.457.3593F}}. PAH heating is potentially the most important heating mechanism \citep[left hand panel in Figure 2 of][]{2016MNRAS.457.3593F}. The PAH abundance in the outer regions of discs is highly uncertain, with some evidence for depletion in the outer disc \citep[e.g.][]{2006A&A...459..545G, 2010ApJ...714..778O, 2011ApJ...727....2P}. The PAH scheme used in \textsc{torus-3dpdr} follows Wolfire et al. (2003) and assumes a PAH--to--dust mass ratio for the ISM of $2.6\times10^{-2}$. In prior studies of external disc photoevaporation \cite{2016MNRAS.457.3593F} assumed this ISM like PAH abundance and  \cite{2017MNRAS.468L.108H}, \cite{2018MNRAS.475.5460H} conservatively assumed a negligible PAH abundance. Similarly \cite{2015ApJ...804...29G} use a low PAH abundance of one hundredth the ISM value. Here we choose an intermediate option. Given the abundance uncertainty and possible depletion at large radii of PAHs in discs, for this initial grid we use a PAH abundance of 10\,per cent the canonical ISM value, i.e. a PAH-to-dust mass ratio of $2.6\times10^{-3}$.

The grid could be expanded in future to account for variations in PAH abundance and/or metallicity. Following \cite{2016MNRAS.457.3593F} at this stage for simplicitys we always assume a mean particle mass of 1.3 in the hydrodynamics.

\begin{table}
 \centering
  \caption{The upper section is a summary of the species included and initial gas abundances for the reduced network used in this paper. The sum of  hydrogen atoms in atomic and molecular hydrogen is unity. The other abundances are with respect to the sum of hydrogen. The lower sections summarise the other microphysical parameters.}
  \label{table:speciesparams}
  \begin{tabular}{@{}l c l c@{}}
   \hline
   \hline    
  \textbf{Gas} \\
  \hline
   Species & Initial abundance & Species & Initial abundance \\
   \hline
   H & $4\times10^{-1}$ & H$_2$ & $3\times10^{-1}$ \\
   He & $8.5\times10^{-2}$ & C+ & $2.692\times10^{-4}$\\
   O & $4.898\times10^{-4}$ & Mg+ & $3.981\times10^{-5}$ \\
   H+ & 0 & H$_2$+ & 0 \\
   H$_3$+ & 0& He+ & 0 \\
   O+ & 0 & O$_2$ & 0 \\
   O$_2$+ & 0 & OH+ & 0 \\
   C & 0 & CO & 0 \\
   CO+ & 0 & OH & 0 \\
   HCO+& 0 & Mg & 0 \\
   H$_2$O& 0 & H$_2$O+ & 0 \\
   H$_3$O & 0 & CH & 0 \\
   CH+ & 0 & CH$_2$ & 0 \\
   CH$_2$+ & 0 & CH$_3$ & 0 \\
   CH$_3$+ & 0 &CH$_4$ & 0 \\ 
   CH$_4$+ & 0 & CH$_5$+ & 0 \\    
   e$^-$  & 0 \\      
   \\      
   \hline    
   \hline    
   \textbf{Dust} \\
   \hline
    $\sigma_{\textrm{FUV}}$ & $2.7\times10^{-23}$\,cm$^{2}$ & \multicolumn{2}{c}{Cross section in wind}  \\
    $\delta$ & $3\times10^{-4}$ & \multicolumn{2}{c}{Dust-to-gas mass ratio in wind}\\
    $f_{\textrm{PAH}}$ & 0.1 & \multicolumn{2}{c}{PAH abundance relative to ISM}\\        
    $\delta_{\textrm{PAH}}$ &$2.6\times10^{-3}$& \multicolumn{2}{c}{PAH-to-dust mass ratio}\\        
   \hline
   \\    
   \hline
   \hline   
   \textbf{Other} \\
   \hline
   $\zeta $ & $5\times10^{-17}$\,s$^{-1}$ & \multicolumn{2}{c}{Cosmic ray ionisation rate}\\
\hline
\end{tabular}
\end{table}

\subsection{Disc construction}
Across our parameter space we are computing the mass loss rate in the photoevaporative wind $\dot{M_w}$ as a function of the stellar mass $M_*$, disc mass $M_d$, disc radius $R_d$ and incident UV field strength $G_0$. For each model we set up a disc structure, which acts as a boundary condition to the photoevaporative wind and is not allowed to dynamically evolve. 

For the disc that sets the boundary condition to the flow we consider a truncated power law surface density profile
\begin{equation}
	\Sigma(R) = \Sigma_{1\textrm{AU}}\left(\frac{R}{\textrm{AU}}\right)^{-1}
	\label{equ:sigma(R)}
\end{equation}
where
\begin{equation}
	\Sigma_{1\textrm{AU}} = \frac{M_d}{2\pi R_d\,1\textrm{AU}}
\end{equation}
Given that our calculations are 1D we require a volume density from a scale height
\begin{equation}
	H = \frac{c_s}{\Omega}
\end{equation}
and
\begin{equation}
	\rho_{\textrm{mid}} = \frac{\Sigma(R)}{\sqrt{2\pi}H}. 
\end{equation}
Unless the disc is externally heated above the temperature set by the central star, the thermal structure in the disc is assumed to be of the form
\begin{equation}
	T = T_{1\textrm{AU}}\left(\frac{R}{AU}\right)^{-1/2}
	\label{equn:discT}
\end{equation}
where $T_{1AU}$ is assumed to vary with stellar mass according to
\begin{equation}
	T_{1\textrm{AU}} = 100\left(\frac{M_*}{M_\odot}\right)^{1/4}\,\textrm{K}. 
\end{equation}
The above is imposed out to some disc outer radius $R_d$ (the values of which are summarised in section \ref{sec:params}). The remainder of the computational domain (see section \ref{sec:otherParams}) freely evolves until it reaches a steady state. Note that the mass loss rate is only sensitive to the disc outer edge, so although we set the conditions there in terms of some global disc properties (e.g. equation \ref{equ:sigma(R)}, and referring to a ``disc mass'') the interior parts of the disc could be different and still give the same mass loss rate if the outer disc were the same. 

The disc itself can also be heated by the external UV field. We apply the maximum of temperatures given by equation \ref{equn:discT} and that computed in the PDR calculation, both within in the disc (which modifies the scale height) and in the flow. External irradiation can hence lower the mid-plane density and increase the scale height of the disc, which in turn can affect the mass loss rate compared to a model where the disc itself is assumed to never be externally heated. 

We also impose a limit on where the photodissociation region (PDR) microphysics is applicable. In 1D models high density, highly optically thick regions can lead to spurious heating in PDR codes as the escape probability of cooling photons tends to zero. Therefore, interior to the point in the flow with an extinction greater than 10 and a gas local number density exceeding $10^{8}$\,cm$^{-3}$ we assume that the temperature of the flow is constant until heating from the star dominates.  This is the approach used by \cite{2016MNRAS.457.3593F} and \cite{2018MNRAS.475.5460H}.

\subsection{Other calculation parameters}
\label{sec:otherParams}
All models in the \textsc{fried} grid use an adaptive mesh, with higher refinement on the simulation domain when and where it is required. For the bulk of our calculations the grid size is 1000\,AU with a maximum resolution of $0.49$\,AU. For the models with $R_d=5$\,AU and $1$\,AU we use grid sizes of 668 and 334\,AU and maximum resolution of 0.33 and 0.16\,AU respectively. 

A unique solution for a steady  transonic flow corresponds to the case
where the flow satisfies criticality conditions \citep[simultaneous
vanishing of terms in the combined momentum and continuity equations, e.g.][]{1965SSRv....4..666P, 2016MNRAS.460.3044C} at some point in the flow. Our numerical scheme will only
converge on this solution if the critical point is contained within the grid.
For a radial flow where temperature is a function of optical depth and local
density, the criticality condition can be written \citep{2016MNRAS.457.3593F}:
\begin{equation}
	\frac{v^2\mu m_{\textrm{H}}}{k_{\textrm{B}}} - T - n \frac{\textrm{d}T}{\textrm{d}n} \geq 0.
	\label{equn:rCrit}
\end{equation}
We therefore retrospectively checked that this was the case, using neighbouring cells for the derivative, and re-ran calculations where necessary with a larger grid. 

\subsection{Scope of the FRIED grid}
\label{sec:params}
Our large grid consists of models that are systematically varied. We include stellar masses of 0.05, 0.1, 0.3, 0.5, 0.8, 1, 1.3, 1.6 1.9 M$_{\odot}$. For each of these stellar masses we model disc masses in a 400\,AU disc that would have been $3.2\times10^{-3}$, 0.1, 1.12, 3.16, 8.94  and  20\,per cent the stellar mass.  We then consider this disc surface density profile truncated at 1, 5, 10, 20, 30, 40, 50, 75, 100, 150, 200, 250, 300, 350, 400\,AU. Finally, for each of these stellar/disc parameters, we consider incident UV fields of 10, $10^2$, $10^3$, $5\times10^3$ and $10^4$\,G$_0$. Note that in practice in a $10^4$\,G$_0$ environment there will also be a strong EUV field, which is not included here. Overall this leads to a substantial grid of 4050 models, though we discard some of the very lowest mass disc models in the 0.05 and 0.1\,M$_\odot$ star cases that appear numerically unstable, leaving us with 3930 models. In this first iteration of the \textsc{fried} grid we consider only a single metallicity and polycyclic aromatic hydrocarbon (PAH) abundance (detailed above in section \ref{sec:num_meth}), though in future we could extend the grid to explore this. 

\subsection{Illustrative calculation}
\begin{figure*}
	\vspace{-0.15cm}
	\includegraphics[width=8.8cm]{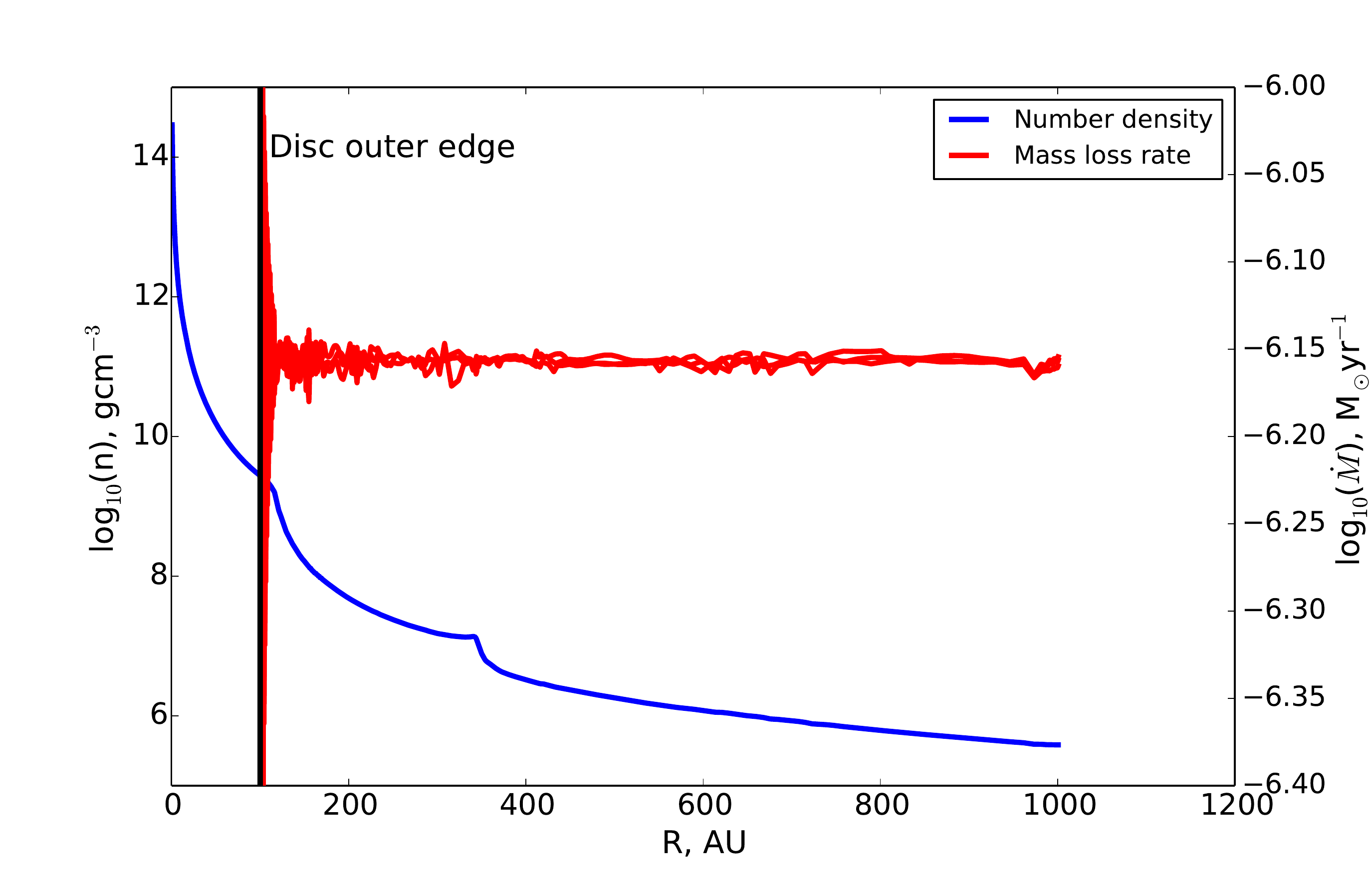}
	\includegraphics[width=8.8cm]{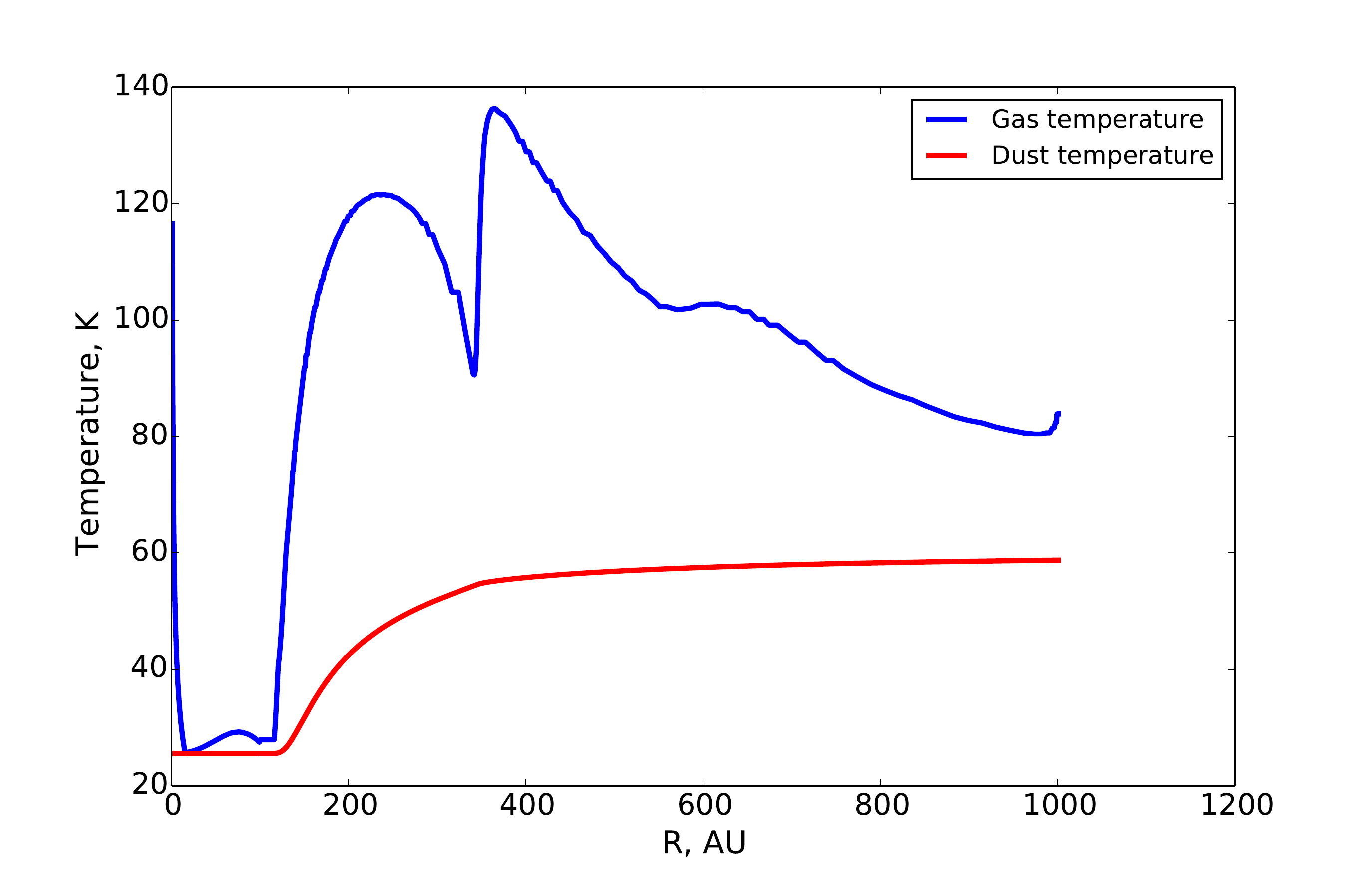}	

	\vspace{-0.15cm}	
	\includegraphics[width=8.8cm]{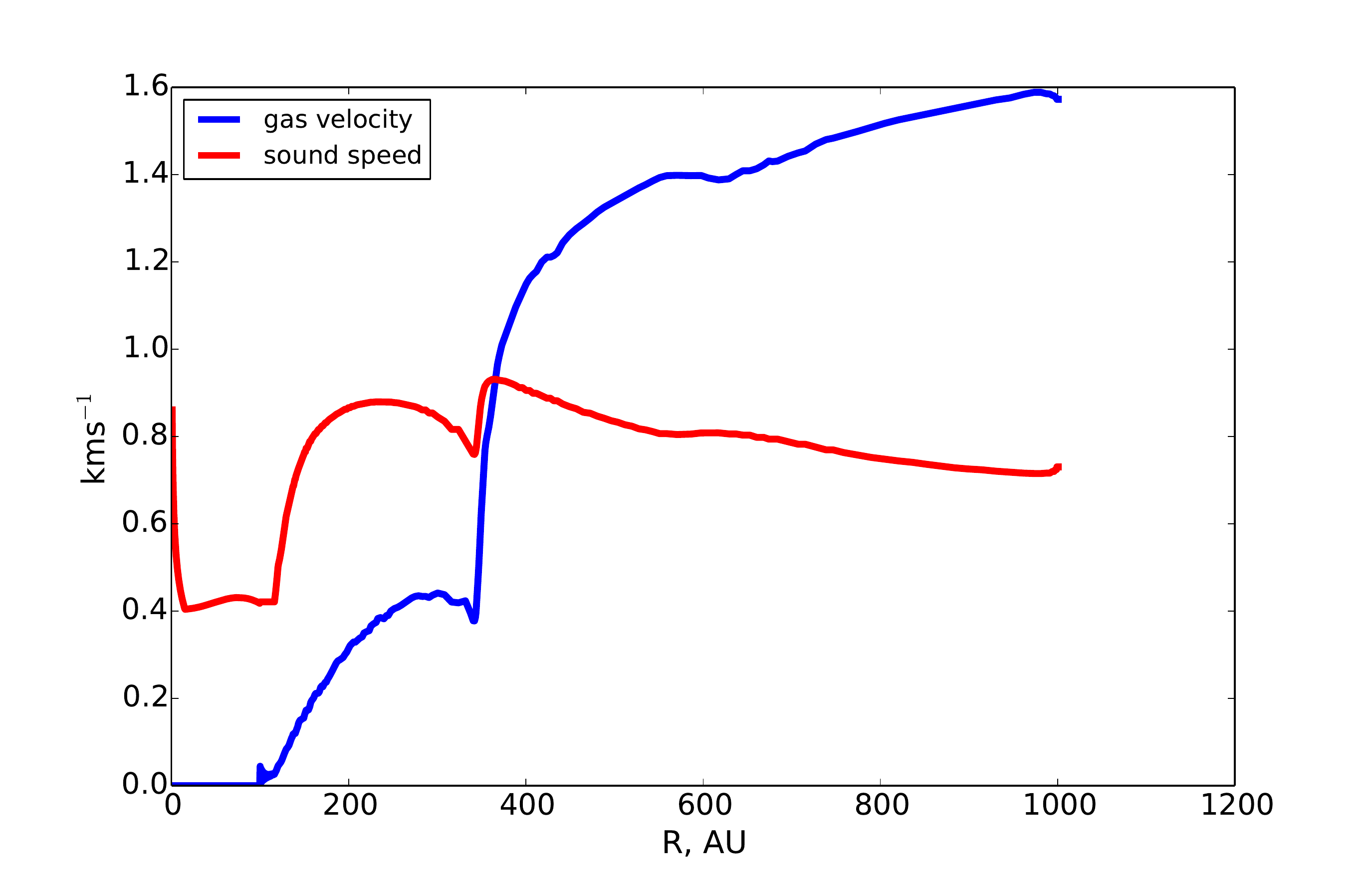}
	\includegraphics[width=8.8cm]{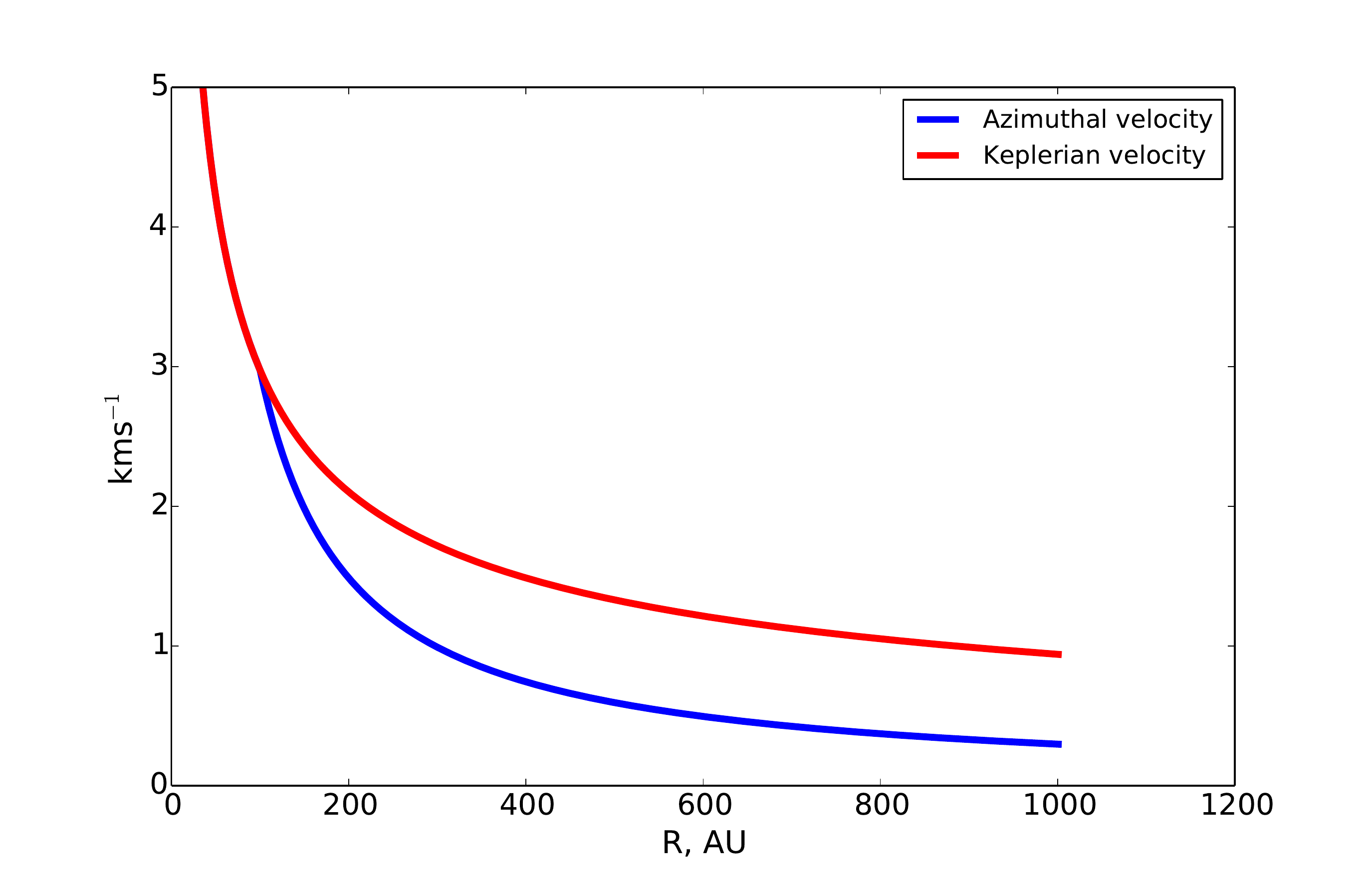}
	
	\vspace{-0.15cm}	
	\includegraphics[width=8.8cm]{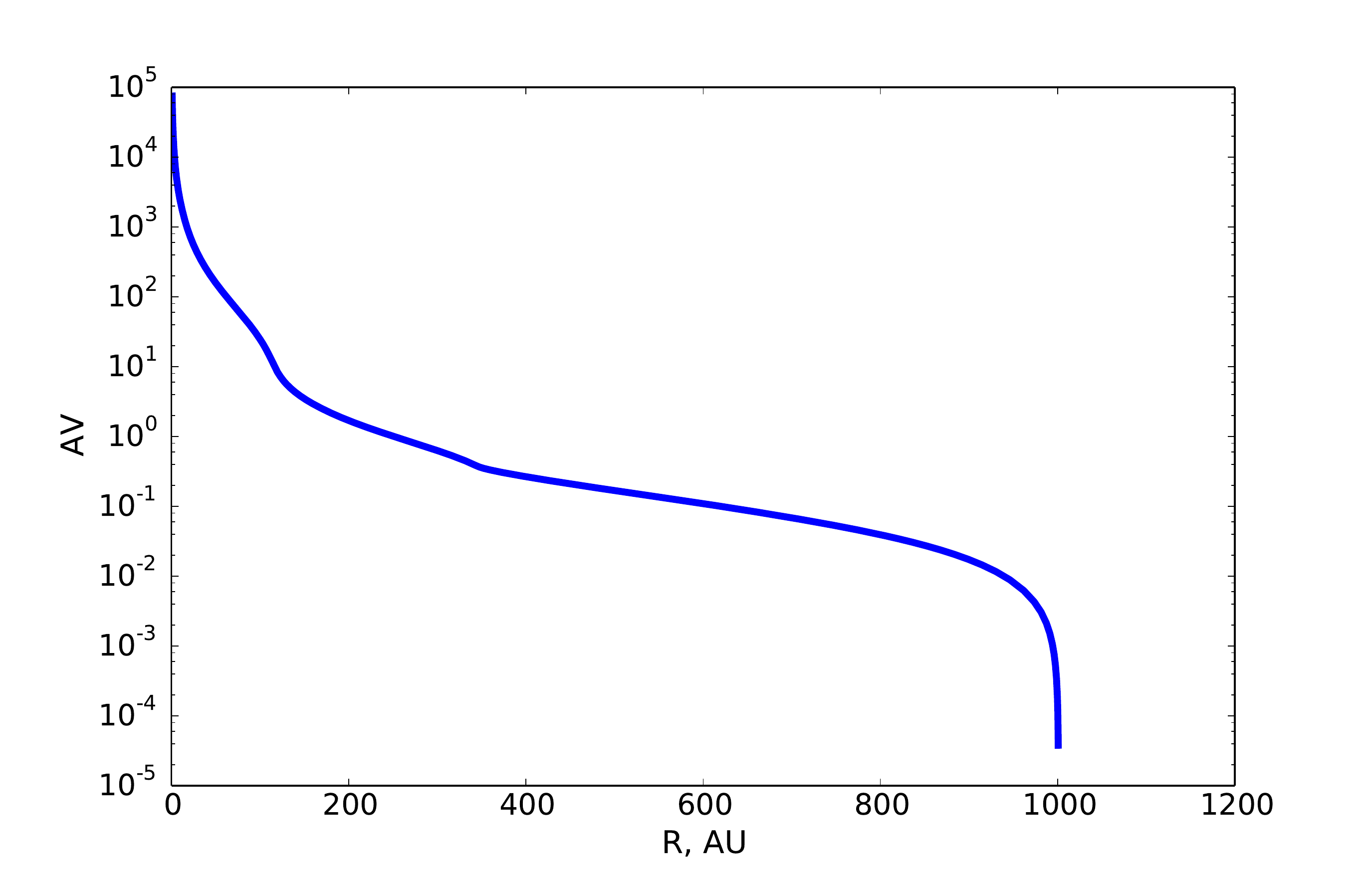}	
	\includegraphics[width=8.8cm]{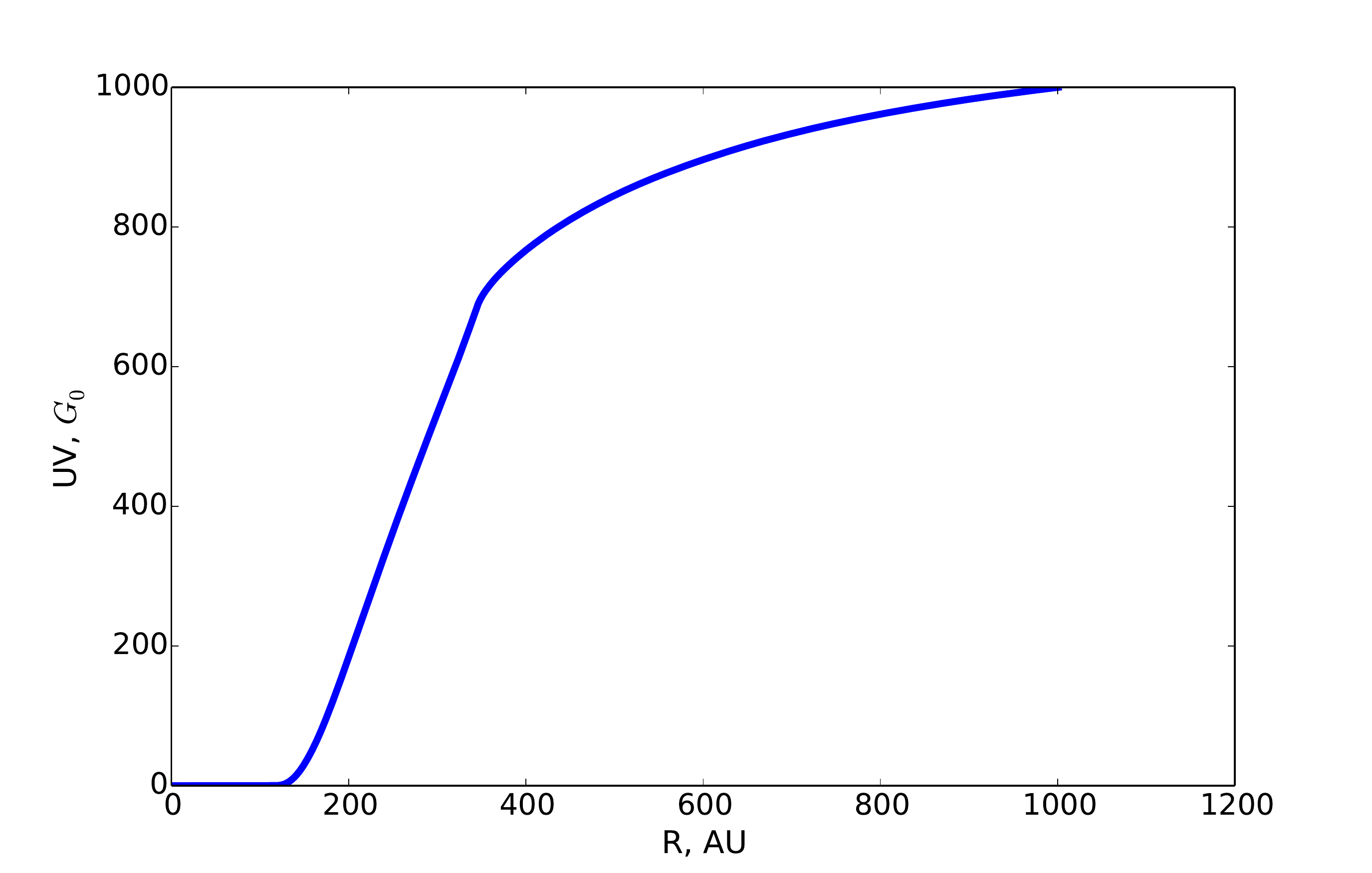}	
	
	\vspace{-0.15cm}	
	\includegraphics[width=8.8cm]{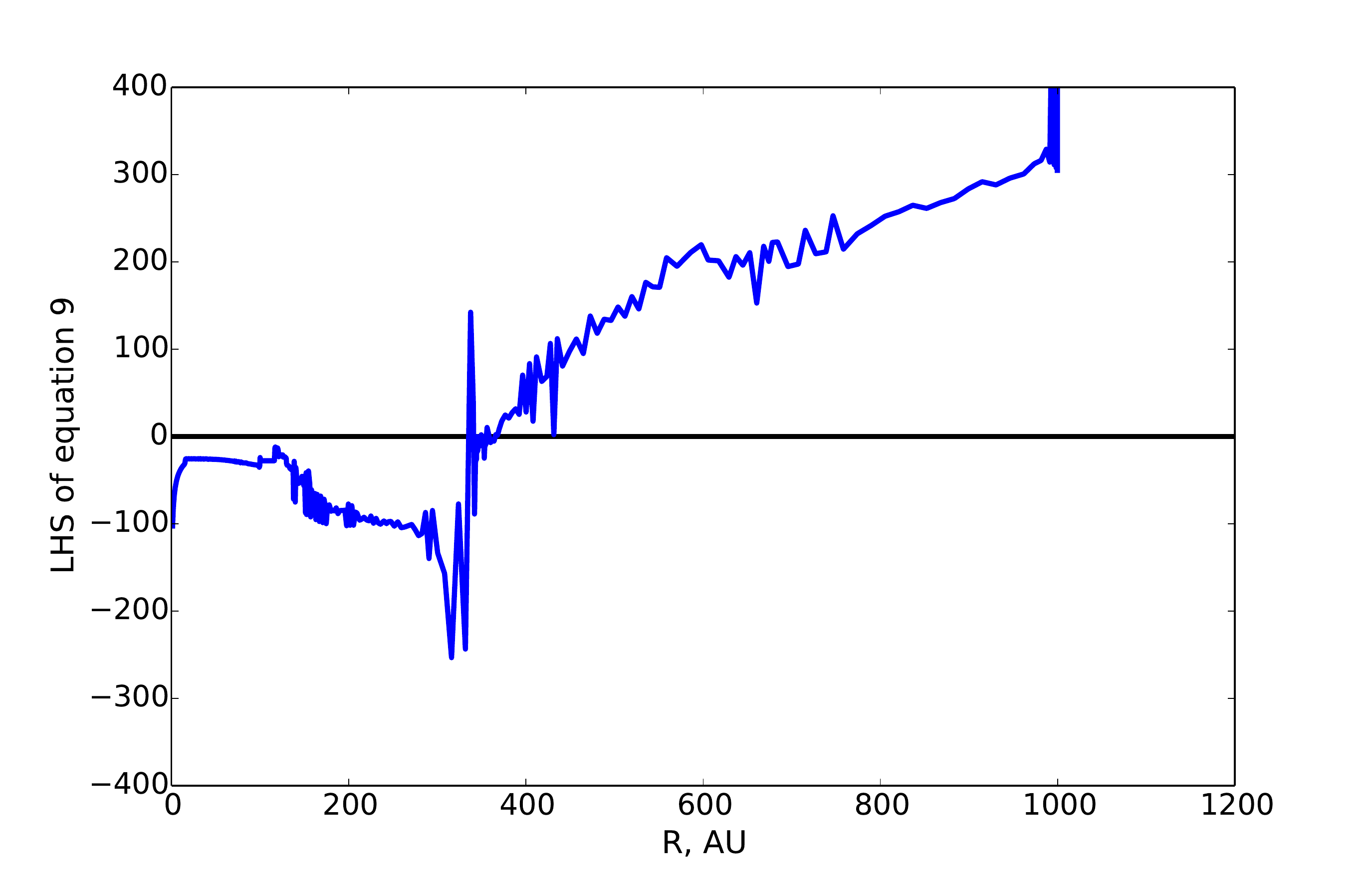}
	\includegraphics[width=8.8cm]{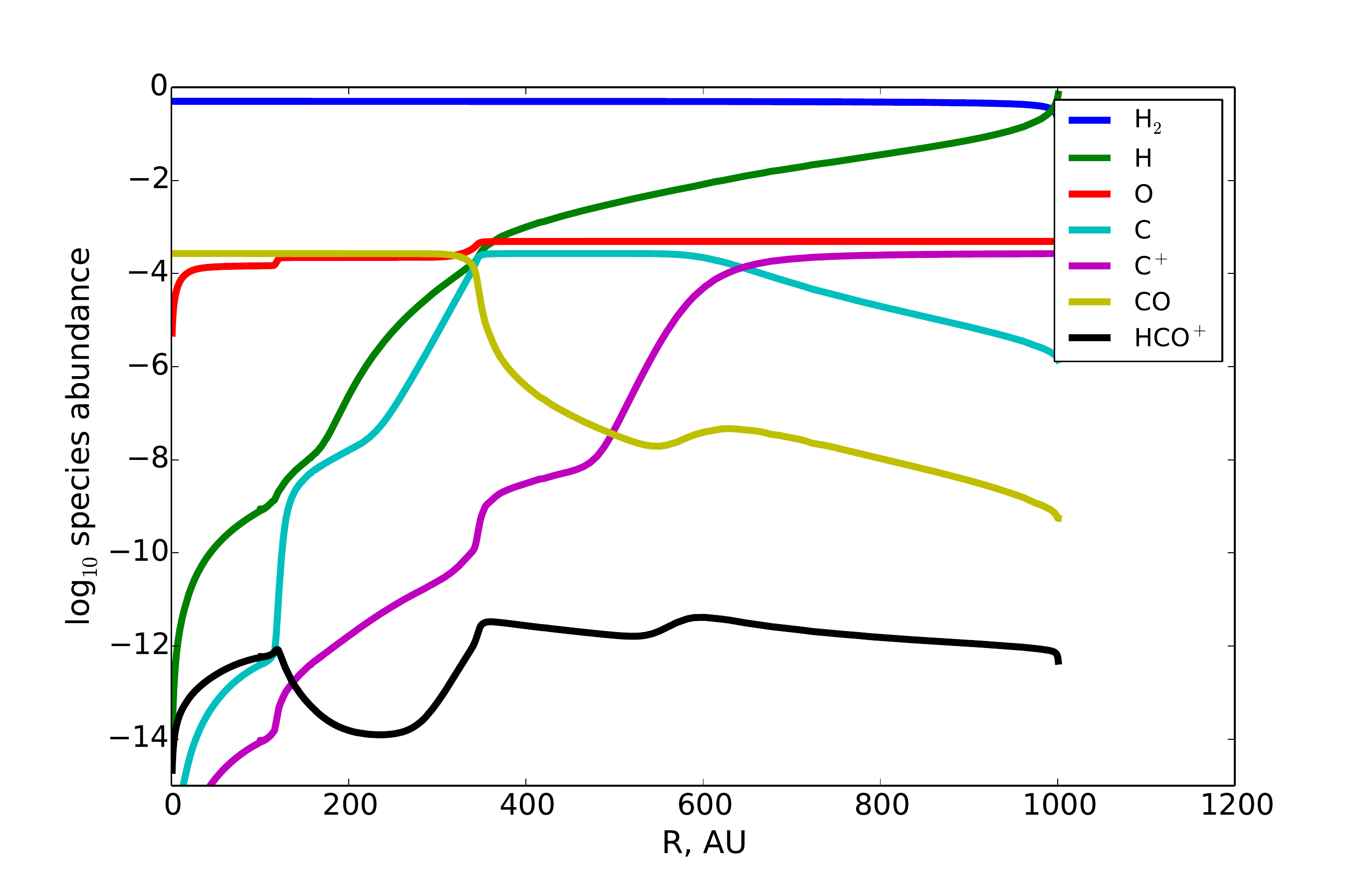}	
	\caption{{A selection of properties of a 100\,AU, 23.4\,M$_{\textsc{jup}}$ disc around a 1\,M$_\odot$ star being irradiated by a $10^3$\,G$_0$ UV field. From left to right, top to bottom these are: the density and mass loss rate (in this panel both are plotted at 3 different times), gas and dust temperature, gas velocity/sound speed, azimuthal and Keplerian velocity, extinction and UV. The left hand lower panel shows the critical point: that at which the left hand side of equation \ref{equn:rCrit} first crosses zero. The lower right hand panel shows the abundance profile of various species }}
	\label{fig:egSingleModel}	
\end{figure*}

{Finally, before presenting the grid itself we illustrate the manner of a single of calculation in detail so the reader can better appreciate the material underlying the grid.  Figure \ref{fig:egSingleModel} shows a series of profiles for a 100\,AU, 23.4\,M$_{\textsc{jup}}$ disc around a 1\,M$_\odot$ star. The top left panel shows the steady state number density profile and mass loss rate profile, with the black vertical line denoting the disc outer edge. In this first panel results from 3 different times are included. There are small variations in the mass loss profile due to numerically induced velocity perturbations induced at the disc outer edge boundary condition, but the mass loss rate is otherwise very steady at $\log_{10}(\dot{M})\sim-6.155$. The upper right panel shows the gas and dust temperature profile. The second row, left panel shows the outward gas velocity and local sound speed profiles. In this case the jump to a supersonic flow coincides with the critical point in the flow, which is illustrated by the crossing of 0 in the lower left hand panel. }

{The second row, right hand panel of Figure \ref{fig:egSingleModel} shows the azimuthal and Keplerian velocity. Deviation from Keplerian velocity is one possible way of detecting external photoevaporation as pointed out by \cite{2016MNRAS.457.3593F} and \cite{2016MNRAS.463.3616H} and possibly identified in the case of IM Lup by  \cite{2018A&A...609A..47P}. The third row left and right hand panels show the extinction and UV profiles and the lower right shows the abundance profile of a small fraction of the 33 species in the network. Note that although our initial abundances in Table \ref{table:speciesparams} are atomic/ionic, once a steady state is achieved the disc (as well as the inner part of the flow itself) is molecular, and molecular line cooling plays an important role in setting the thermal balance there.}

{This makes up just one of the 3930 models of the \textsc{fried} grid which are all similarly complex. }

\section{The fried grid}
\label{sec:theFriedGrid}

\subsection{Mass loss rates}
The mass loss rates over the entire grid are summarised in Figures \ref{fig:all10G0}--\ref{fig:all10000G0}. Each figure consists of a series of panels plotting the mass loss rate as a function of the disc size and mass. The different panels in each figure correspond to different stellar masses, as indicated  above each panel. We do not intend to describe each component of the grid in detail, but note the following highlights before considering some immediate consequences in section \ref{sec:discussion}.

\begin{figure*}
	\includegraphics[width=8.5cm]{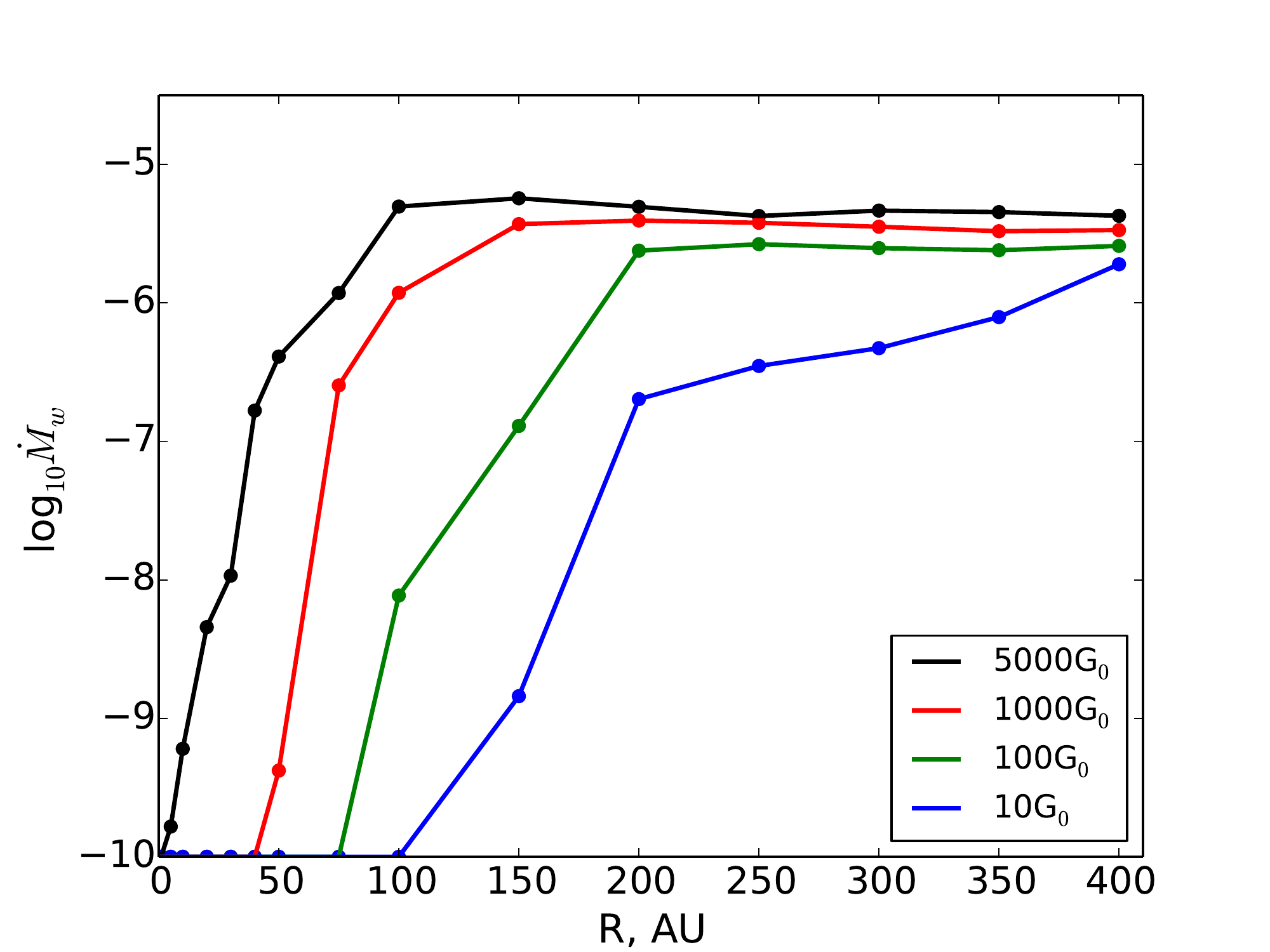}
	\includegraphics[width=8.5cm]{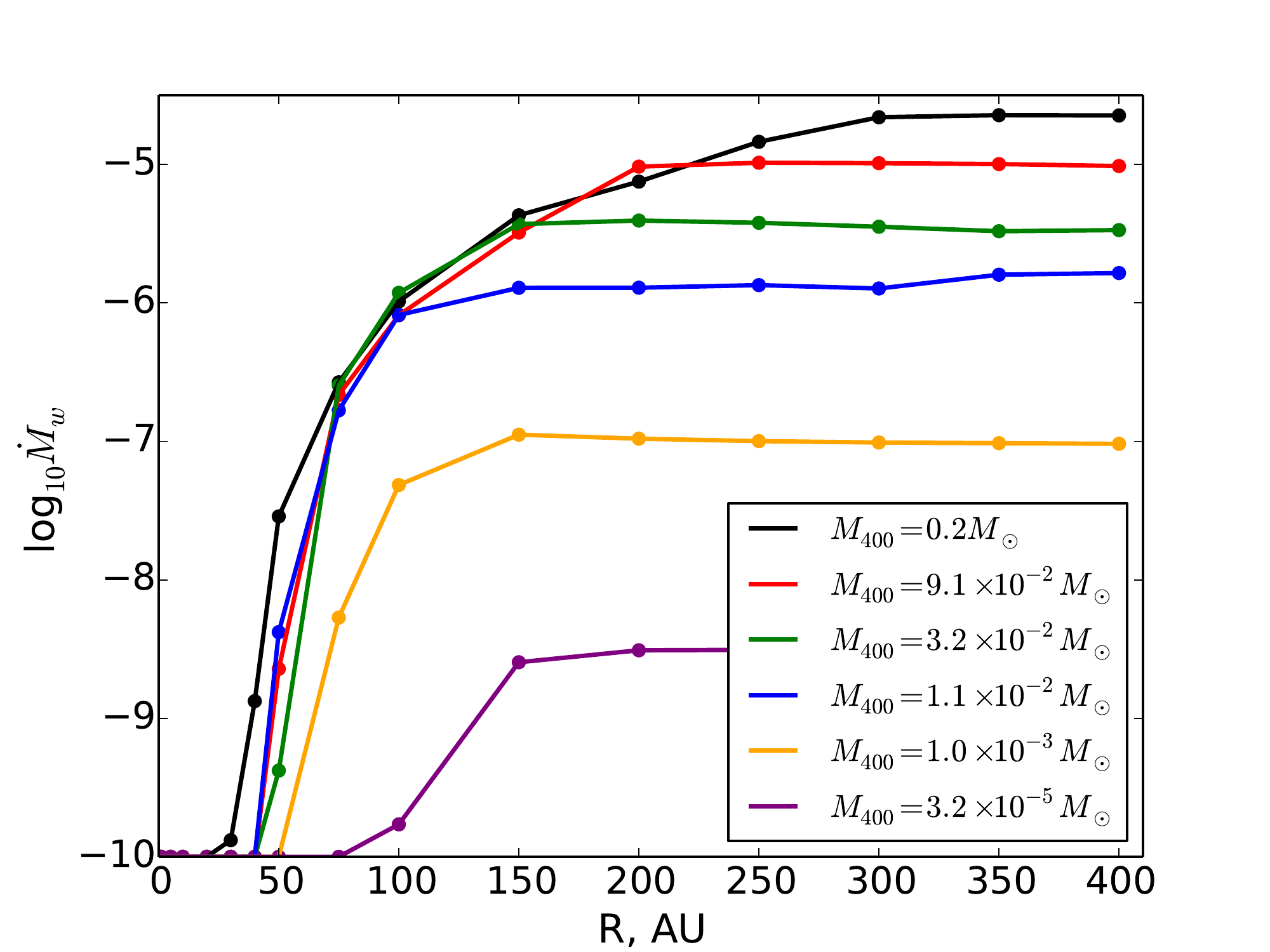}
	\caption{Illustrative mass loss rates as a function of disc outer radius. The left hand panel varies the UV field strength incident upon a $2.2\times10^{-2}$\,M$_\odot$, 100\,AU disc. The right hand panel varies the disc mass out to 400\,AU for a fixed UV field of 1000\,G$_0$. All of the above are for a 1\,M$_\odot$ star}
	\label{fig:egProfiles}	
\end{figure*}	

\begin{enumerate}
	\item An illustration of mass loss rates as a function of radius is given in Figure \ref{fig:egProfiles}. The left hand panel holds the star--disc parameters constant, but varies the incident UV field strength. The right hand panel is for a $10^3$\,G$_0$ field, but different disc masses. The disc size at which external evaporation is ineffective increases with decreasing UV field strength. For a fixed UV field strength the mass loss rate increases with the disc mass, but the radius at which external photoevaporation becomes ineffective is relatively constant. We struggle to find steady state mass loss rates below $10^{-10}$\,M$_\odot$\,yr$^{-1}$, so impose this as the floor value of the grid.  This is lower than the internal mass loss rate due to EUV/X-rays \citep[e.g.][]{2012MNRAS.422.1880O}. \\

	\item For UV fields $< 10^4$\,G$_0$, as the stellar mass increases photoevaporation becomes ineffective out to larger and larger disc radii. For example in the 10\,G$_0$ case, for a 1\,M$_\odot$ star photoevaporation is ineffective for discs $<100$\,AU in size (see the middle right panel of Figure \ref{fig:all10G0}), but in the 0.1\,M$_\odot$  case photoevaporation is effective down to $\sim10$\,AU (top middle pane of Figure \ref{fig:all10G0}). \\

	\item In the $10^4$\,G$_0$ case (Figure  \ref{fig:all10000G0}) photoevaporation is effective at all radii for all stellar 
masses that we consider (up to 1.9\,M$_\odot$). This is because the UV field is high enough that it can still heat the disc to a sufficient extent that a wind can be driven, even when it is very compact. This implies that an infrared excess due to small grains would disappear very quickly for discs in regions such as the Orion Nebular Cluster (ONC) especially given the additional effects of EUV
heating (not included here) in this environment. \\ 

	\item In the lowest stellar mass cases (0.05 and 0.1\,M$_\odot$) the mass loss rate is relatively insensitive to the UV field strength for extended
discs ($R_d>\sim20$\,AU). For example for a 100\,AU, 0.15\,M$_{\textrm{jup}}$ disc  around a 0.05\,M$_\odot$ star the mass loss rate increases by only around 1\,per cent from 10 to $10^4$\,G$_0$. The temperature in the flow does steadily increase with UV field strength, resulting a more rarefied, but also faster flow. This lower density in the flow coupled with higher velocity as the UV field increases means that the mass loss rate stays approximately constant. \\
	
	\item Discs around low mass stars ($\leq0.3$\,M$_\odot$) are evaporated down to small radii ($<50$\,AU) much more effectively than higher mass stellar systems. This is because the disc is less gravitationally bound to a lower mass parent star. \\
\end{enumerate}

\subsection{Behaviour as $R_{\textrm{H}-\textrm{H}_2}\rightarrow R_d$}
Despite the $10^4$\,G$_0$ case giving significant mass loss down to the smallest disc outer radii (see Figure \ref{fig:all10000G0}), it can exhibit slightly lower mass loss rates than in the  $(5\times)10^3$\,G$_0$ cases at larger disc outer radii. We explored this using some additional models for a 100\,AU disc around a 1\,M$_\odot$ star with incident UV fields from $2\times10^3-1.5\times10^4$\,G$_0$. 

\begin{figure}
	\hspace{-0.8cm}
	\includegraphics[width=10.8cm]{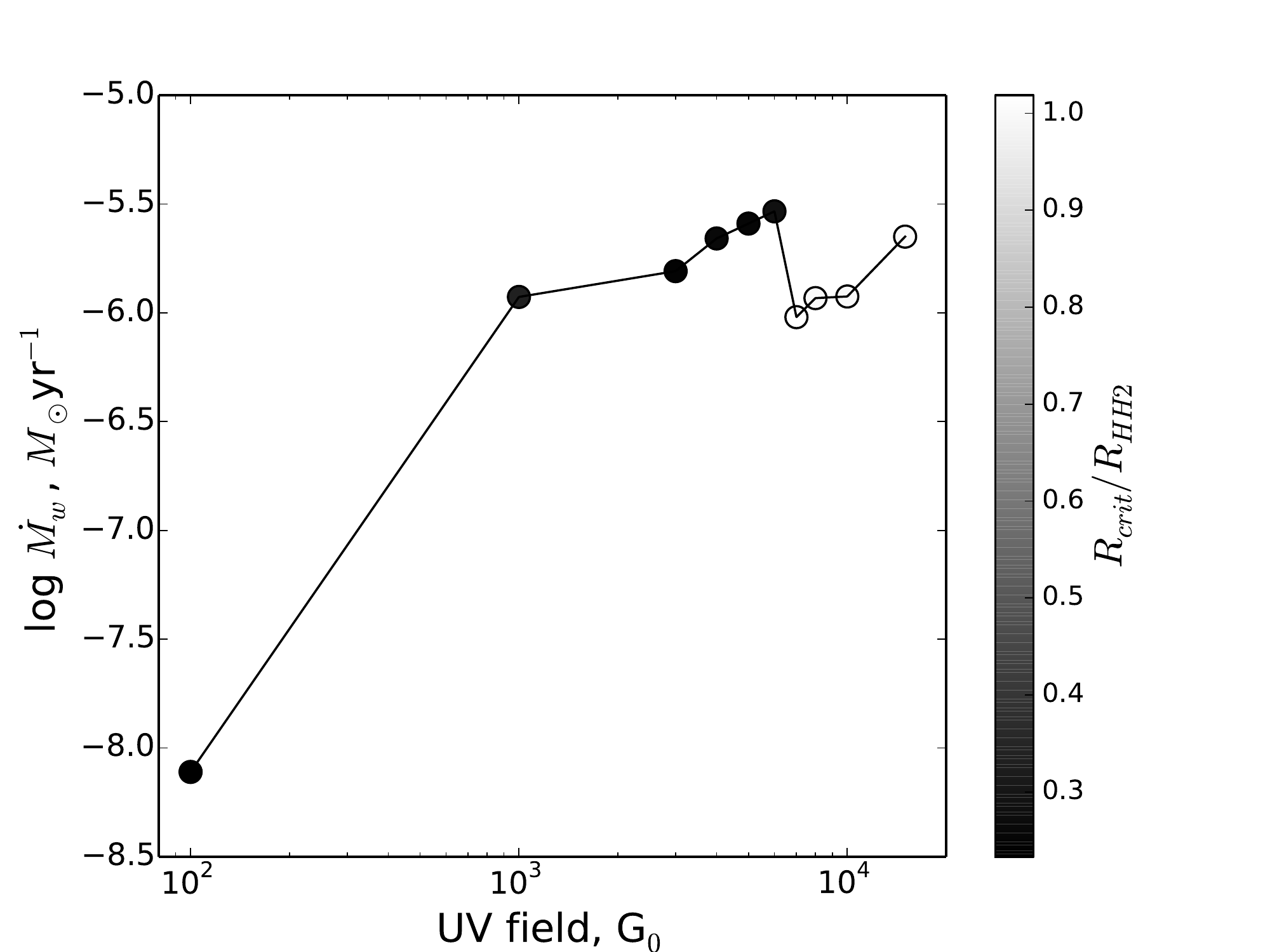}
	\caption{The mass loss rate as a function of UV field strength for the model that we explore in more detail to probe the regime where the H--H$_2$ transition influences the mass loss rate. The colour scale denotes the ratio of the critical radius to the radius of the H--H$_2$ transition. Their coincidence results in a drop in the mass loss rate. }
	\label{fig:HH2Ram}
\end{figure}

The \cite{2016MNRAS.457.3593F} critical point (equation \ref{equn:rCrit}) is usually in the molecular part of the wind, with the H--H$_2$ transition at some larger radius. Comparing these two locations, we find that the drop in mass loss rate at high UV is associated with the H--H$_2$ transition becoming coincident with the critical point. In Figure \ref{fig:HH2Ram} we show the mass loss rate as a function of UV field strength for the additional exploratory models, with points colour coded by the ratio of the critical radius to that of the H--H$_2$ transition. This clearly illustrates that the mass loss rate drop at $\sim7000$\,G$_0$ is associated with the critical radius being coincident with the H--H$_2$ transition. 

We will present a physical description of the flow in this new regime in a subsequent paper, but note now that it shows some similarities to prior analytic work in which the hydrogen ionisation front sets the mass loss rate \citep[e.g.][]{1998ApJ...499..758J}, only with added complications such as the fact that the change in sound speed across the H--H$_2$ transition is unknown.

\subsection{Online resources}
\label{sec:onlinetool}
The raw mass loss rate grids are available for download both as online material accompanying this journal article and from the \textsc{fried} website\footnote{\url{http://www.friedgrid.com/Downloads/}}. In addition to this, we have developed a web interface that allows users to quickly and easily estimate mass loss rates for arbitrary disc/UV field parameters\footnote{\url{http://www.friedgrid.com/Tool/}}. In its current form the user specifies a disc extent, disc mass and stellar mass. Values of the mass loss rate for a range of UV field strengths are then returned using a linear interpolation over the grid with the \textsc{python} \textsc{scipy} \textsc{LinearNDInterpolator} routine. We choose a linear interpolation because it is straightforward and robust, unlike higher order schemes which could possibly give rise to oscillations. Of course by downloading the entire grid a more sophisticated interpolation can be applied at the users own discretion and risk. The online tool does not permit queries to be made beyond the bounds of the grid parameter space. For special cases contact the authors to discuss either expanding the grid or running bespoke models for a given application. 

The intention is that the online tool can be easily be used by theorists and observers alike to quickly gauge whether significant photoevaporation is occuring in a given scenario. An illustrative example of application of the \textsc{fried} grid to real systems is given in section \ref{sec:illustrativeReal}.

\section{Discussion}
\label{sec:discussion}
\begin{figure*}
	\centering
	\includegraphics[width=19cm]{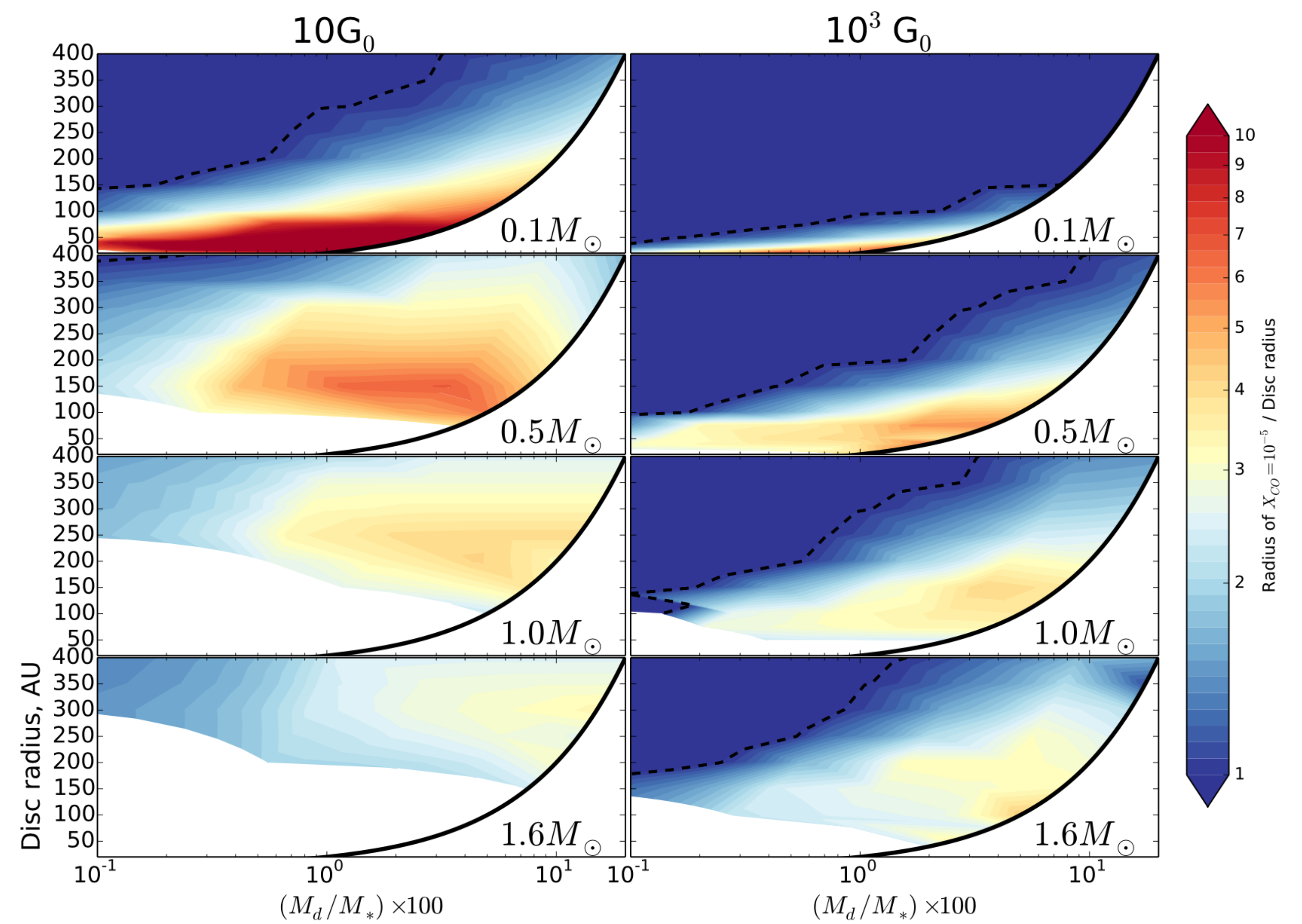}
	\caption{The ratio of the extent out to which CO is abundant to the extent of the Keplerian disc, as a function of disc size and the ratio of disc--to--stellar mass. The dashed contours denote the point at which CO is depleted at the disc outer edge. Models with mass loss rates less than $10^{-9}$\,M$_\odot$\,yr$^{-1}$ are not included and the region below the solid black line is not included in our parameter space.  The left and right hand panels are for UV fields of 10 and $10^3$\,G$_0$ respectively and the stellar mass 0.1, 0.5, 1 and 1.6\,$M_\odot$ from top to bottom.  If CO extends beyond the disc outer edge, then it is possible it could be used to detect the photoevaporative wind, for example through a sub-Keplerian rotation profile or a radially increasing temperature measured with CO line ratios. {Note that since the grid extent is typically 1000\,AU, some of the ratios presented here may be lower limits if the grid size is what limits the CO extent.  }}
	\label{fig:COextents}
\end{figure*}

\subsection{Constraints on detecting external photoevaporation using CO in weak--intermediate UV regimes}
The PDR thermal physics in our models also contains information on the chemical structure of the flow which could provide clues as to the best scenarios in which to search for external photoevaporation in action. CO observations towards discs are common, but typically focus on rings and gaps in the main body of the disc, rather than the outer disc where external photoevaporation is expected to be strongest. Two possible signatures of external photoevaporation in CO are sub-Keplerian rotation in the outer disc \citep{2016MNRAS.457.3593F, 2016MNRAS.463.3616H} and a radially increasing temperature profile. Sub-Keplerian rotation in the candidate evaporating disc IM Lup \citep{2016ApJ...832..110C, 2017MNRAS.468L.108H} has been detected by \cite{2018A&A...609A..47P}. A radially increasing temperature profile could be probed using CO line ratios.  

Given the above and current lack of other diagnostics, of specific interest are regions of the parameter space where a kinematic and thermal tracer such as CO is observable and the mass loss rate is non-negligible (and, ideally, the deviation from Keplerian rotation is spectrally resolvable). {We therefore compute the ratio of the radius at which CO is reduced in abundance ($R_{\textrm{CO}}$)  to the disc outer radius ($R_{\textrm{disc}}$). This transition happens quickly, so we mark the point at which the CO abundance drops to less than $10^{-5}$. Note that the role of different reactions as well as cosmic rays and photodissociation on setting the CO abundance are well known and the reader is directed to \cite{1986ApJS...62..109V} and \cite{1988ApJ...334..771V} for more information.} Values of  $R_{\textrm{CO}}/R_{\textrm{disc}}>1$ therefore have CO abundant in the photoevaporative wind. Of these models, we then only retain those in which the mass loss rate is greater than $10^{-9}$\,M$_{\odot}$\,yr$^{-1}$. This leaves us with a grid of models that summarise the best regimes in which it might be possible to probe a photoevaporative wind using CO.

Figure \ref{fig:COextents} summarises this dataset for a 0.1, 0.5, 1 and 1.6\,M$_\odot$ stars in 10 and $10^3$\,G$_0$ environments. In the low UV case if the stellar mass is low then CO is abundant in the flow and there is still significant mass loss down to small disc outer radii ($<50$\,AU). However at higher stellar masses there is a lower limit on the radius of the disc at which the mass loss rate is actually significant. At higher UV field strengths in the low stellar mass scenario CO is depleted in the photoevaporative wind except for very compact discs. Conversely for higher stellar masses CO is abundant \textit{and} has significant mass loss down to smaller disc outer radii than in the 10\,G$_0$ case. Discs around $1-2$\,M$_\odot$ stars in $\sim10^3$\,G$_0$ environments therefore seem to provide a good opportunity to identify external photoevaporation in action.

Another interesting point to note is that the typical factor 2 or more extent of CO relative to the Keplerian disc outer edge is large enough to account for the observed extent of the gas relative to dust in Lupus discs by \cite{2018ApJ...859...21A}. Since only small grains are entrained in a photoevaporative wind \citep{2016MNRAS.457.3593F} and hence will not be detected in the millimetre continuum this raises the possibility that external photoevaporation might be a contributing factor in setting the observed relative gas--dust disc sizes. However this has the caveat that we cannot robustly constrain the surface brightness in CO from 1D models so cannot currently predict the observable CO extent, just that out to which it is abundant.

\subsection{Illustrative instantaneous mass loss rate estimates for real systems: Taurus discs}
\label{sec:illustrativeReal}
The \textsc{fried} grid makes it trivial to estimate mass loss rates for real systems based on estimates of the stellar mass, disc mass, disc radius and incident UV field strength. In cases where not all of these are known, the grid can be used to assess ranges of plausible values.

\begin{table*}
	\caption{Illustrative mass loss rates for real systems for a range of UV field strengths. All star-disc parameters are from Guilloteau et al. (2011). }
	\label{tab:realApp}
	\begin{tabular}{c c c c c c c c c }
	\hline
	System & $M_* (M_{\odot})$ & $M_d (10^{-3}\,M_\odot)$ & $R_d$ (AU) & $\dot{M} (M_\odot\,\textrm{yr}^{-1})$ & CO in flow?&  High/Medium/Low  \\
	& & & & $10G_0$ & (Figure 1) & Mass loss rate  \\
	\hline
	BP Tau   & 0.78 &   5.4  &    57. & $10^{-10}$ & -- & L  \\
      CI Tau   & 0.76 &  37          &            201. & $4.6\times10^{-7}$  & Y &H  \\
       CQ Tau  &  1.7 &   6.3         &            188.0 & $2.6\times10^{-10}$ & -- & L  \\ 
       CY Tau  &  0.48 &  16.5       &    92. & $9.9\times10^{-8}$ & Y  & H   \\            
       DG Tau  &  0.7  &  36.0   & 198. & $5.2\times10^{-7}$ & Y &H	\\ 
       DL Tau   & 0.7 &   49.0   &  179. & $1.8\times10^{-7}$ &  Y  &H	\\ 
       DM Tau  &  0.47 &  31.1 & 274. & $2.8\times10^{-6}$ & Y &H   \\            
       DQ Tau &   0.55 &  12.1 & 439. & $3.6\times10^{-7}$  & --  & H     \\              
       GM Aur  &  1.37 &  27.0   &  578. & $8.9\times10^{-7}$ & --  &H \\ 
       Lk Ca 15  & 1.12 &  28.4 & 178. & $2.3\times10^{-8}$ & Y &H \\ 
      MWC480 &  1.8 &   182.3  & 155. & ? & -- &?  \\ 
      MWC758  & 1.8  &  10.6   & 187. & $2.3\times10^{-9}$ & -- &M \\ 
      HL Tau  &  0.7 &   90.6  & 280.16 & $3.7\times10^{-6}$  & Y  &H    \\ 
       HH 30  &   0.25 &  8.1 &  123. & $7.5\times10^{-7}$ & Y & H \\ 
     DG Tau B &  3.0 &   67.9   & 303. & ?& -- &  ? \\ 
      T Tau N  &  1.9  &  0.1    &   67. &$10^{-10}$ & -- & L  \\ 
      Haro6-13 & 0.55   & 0.6  &  90. & $4.9\times10^{-10}$ & -- & L    \\            
      Haro6-33 & 0.55 &  0.5  & 439. & $2.6\times10^{-8}$ & -- &  H    \\            
	\hline
	\end{tabular}
\end{table*}
\label{lastpage}

In Table \ref{tab:realApp} we calculate the mass loss rate for the dust disc extents and masses in Taurus inferred by \cite{2011A&A...529A.105G}. PDR modelling of Taurus requires an average UV field of around 10\,G$_0$ according to \cite{2013MNRAS.429.3584H} to give consistent H\,\textsc{i} and CO observations, so we adopt this UV field strength for our illustrative assessment. For discs larger than 400\,AU (the upper limit on our grid) we set the disc radius to 400\,AU, which is, if anything, an underestimate of the mass loss rate (see Figure 1). We do not similarly limit the disc mass. These estimates are based on dust disc radii, so given the gas is generally found to be more extended \citep[e.g.][]{2013A&A...557A.133D, 2017A&A...605A..16F} and mass loss rates are higher for larger discs it is quite a conservative estimate. 

In the sixth column of Table \ref{tab:realApp}  we note whether each disc has CO in the flow from Figure \ref{fig:COextents}. In the final column of Table \ref{tab:realApp} we denote any system in which the mass loss rate is above $10^{-8}$\,M$_\odot$\,yr$^{-1}$ a high mass loss rate (H), otherwise if the mass loss rate is above $10^{-9}$\,M$_\odot$\,yr$^{-1}$ we denote it ``medium'' (M), else the mass loss rate is low (L). In two cases, MWC\,480 and DG Tau B, the disc mass is extremely high, so the \textsc{fried} grid is unable to compute a reliable mass loss rate estimate. 

At least half of the discs in our illustrative application have significant mass loss rates ($>10^{-8}$\,M$_\odot$\,yr$^{-1}$) in a 10\,G$_0$ environment.  Of these discs with high mass loss rates there are 8 objects in which
CO is predicted to exist in the flow over a significant radial
range and therefore make possible targets for probing subtle sub-Keplerian rotation and/or radially increasing temperature profile in the outer disc (there may be other signatures of external photoevaporation that are yet to be identified).



\section{Summary}

We present the \textsc{fried} grid of 3930 models of externally photoevaporating protoplanetary discs. It spans stellar masses from 0.05--1.9\,M$_\odot$, UV fields from $10-10^4$\,G$_0$, disc radii from 1--400\,AU and a range of disc masses. 

Such calculations are technically difficult to compute and computationally expensive, so this grid finally makes consideration of the effects of radiation environment accessible to the community. The entire grid can be downloaded as supplementary data attached to this paper, or from the \textsc{fried} website. Additionally,  we have developed an associated web tool which permits easy computation of the instantaneous mass loss rates for given disc parameters. 

We illustrate application of the grid to estimate mass loss rates for real systems. In addition to providing a large dataset to the community the \textsc{fried} grid also yields the following immediate points of interest. \\

\noindent 1) For stellar masses $\leq0.3$\,M$_\odot$ external photoevaporation is effective down to small disc radii ($<50$\,AU) even for UV fields of 10\,G$_0$.  For higher stellar masses external photoevaporation is only effective down to some radius that is a function of the irradiating UV field strength and the disc mass. Interior to this radius gravity dominates and the mass loss rate is negligible. \\ 

\noindent 2) In a $10^4$\,G$_0$ environment external photoevaporation is effective for all stellar masses at least up to $1.9$\,M$_\odot$ at all disc radii. However, at these high UV field strengths ($\sim7-1.5\times10^4$\,G$_0$) if the H--H$_2$ transition draws close enough to the disc outer edge such that it becomes coincident with the \cite{2016MNRAS.457.3593F} critical point, the H--H$_2$ transition can affect the dynamics and lower the mass loss rate. We will present a physical description of this in an accompanying paper. 
 \\

\noindent 3) We compute illustrative mass loss rates for the \cite{2011A&A...529A.105G} disc parameters in Taurus, finding that around half of those discs are expected to be undergoing significant photoevaporation in a 10\,G$_0$ field \citep[the field strength argued for in Taurus by][]{2013MNRAS.429.3584H}. These estimates are based on dust disc radii which, since the gas is more extended, make the mass loss rate estimates conservative.  \\

\noindent 4) Of the \cite{2011A&A...529A.105G} discs with significant mass loss, most are expected to have CO survive in the photoevaporative outflow, making them possible candidates to try and identify external photoevaporation in action in weak/intermediate UV environments.   \\

\noindent 5) Generally, $1-2$\,M$_\odot$ stars in $\sim10^3$\,G$_0$ environments have high mass loss rates over a large range of radii whilst retaining CO in the flow, making them good candidates for detecting external photoevaporation in action. Conversely in 10\,G$_0$ UV environments the disc is too bound in this stellar mass range for significant photoevaporation once the disc is below around 100\,AU in size.  \\

\section*{Acknowledgements}
{We thank the referee for their positive review and excellent suggestions for improving the paper. }
TJH is funded by an Imperial College London Junior Research Fellowship. This work has been partially
supported  by  the DISCSIM project, grant agreement 341137 funded by the European Research Council under ERC-2013-ADG. The simulations in this paper were primarily computed in the final months and days of the life of the COSMOS Shared Memory system at DAMTP, University of Cambridge operated on behalf of the STFC DiRAC HPC Facility. This equipment is funded by BIS National E-infrastructure capital grant ST/J005673/1 and STFC grants ST/H008586/1, ST/K00333X/1. Part of this work used the DiRAC Data Analytics system at the
University of Cambridge, operated by the University of Cambridge
High Performance Computing Serve on behalf of the STFC DiRAC
HPC Facility (www.dirac.ac.uk). This equipment was funded
by BIS National E-infrastructure capital grant (ST/K001590/1),
STFC capital grants ST/H008861/1 and \\ ST/H00887X/1, and STFC
DiRAC Operations grant ST/K00333X/1.

\bibliographystyle{mnras}
\bibliography{molecular}

\appendix

\section{Mass loss rate grids}

\begin{figure*}
	\centering	
	\includegraphics[width=19cm]{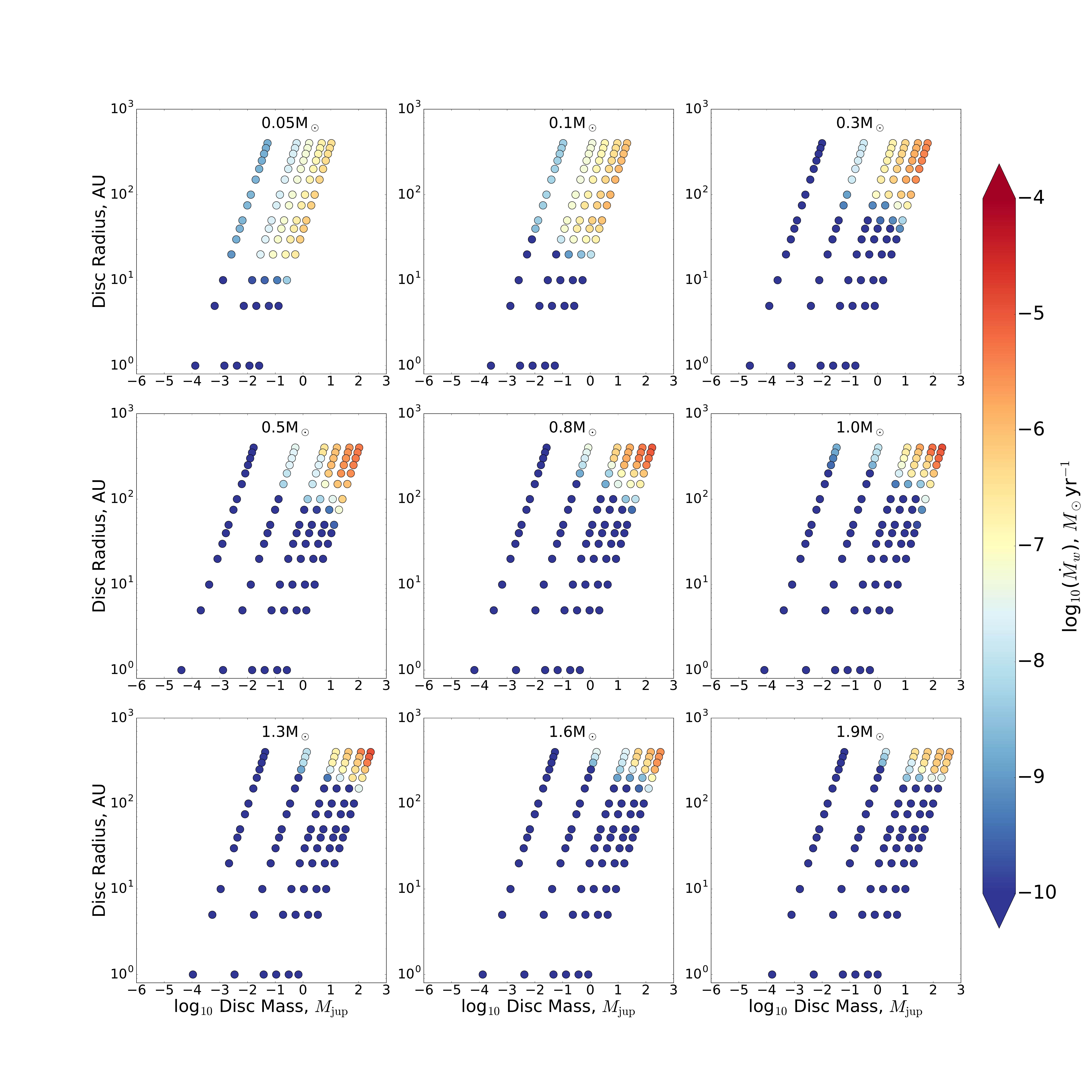}
	\caption{The mass loss rate for the 10\,G$_0$ models over our entire parameter space, plotted as a function of disc mass and outer radius. The stellar mass in each case is denoted above each panel. Note that the grid has a floor value on the mass loss rate of $10^{-10}$\,M$_\odot$\,yr$^{-1}$}
	\label{fig:all10G0}
\end{figure*}

\begin{figure*}
	\centering	
	\includegraphics[width=19cm]{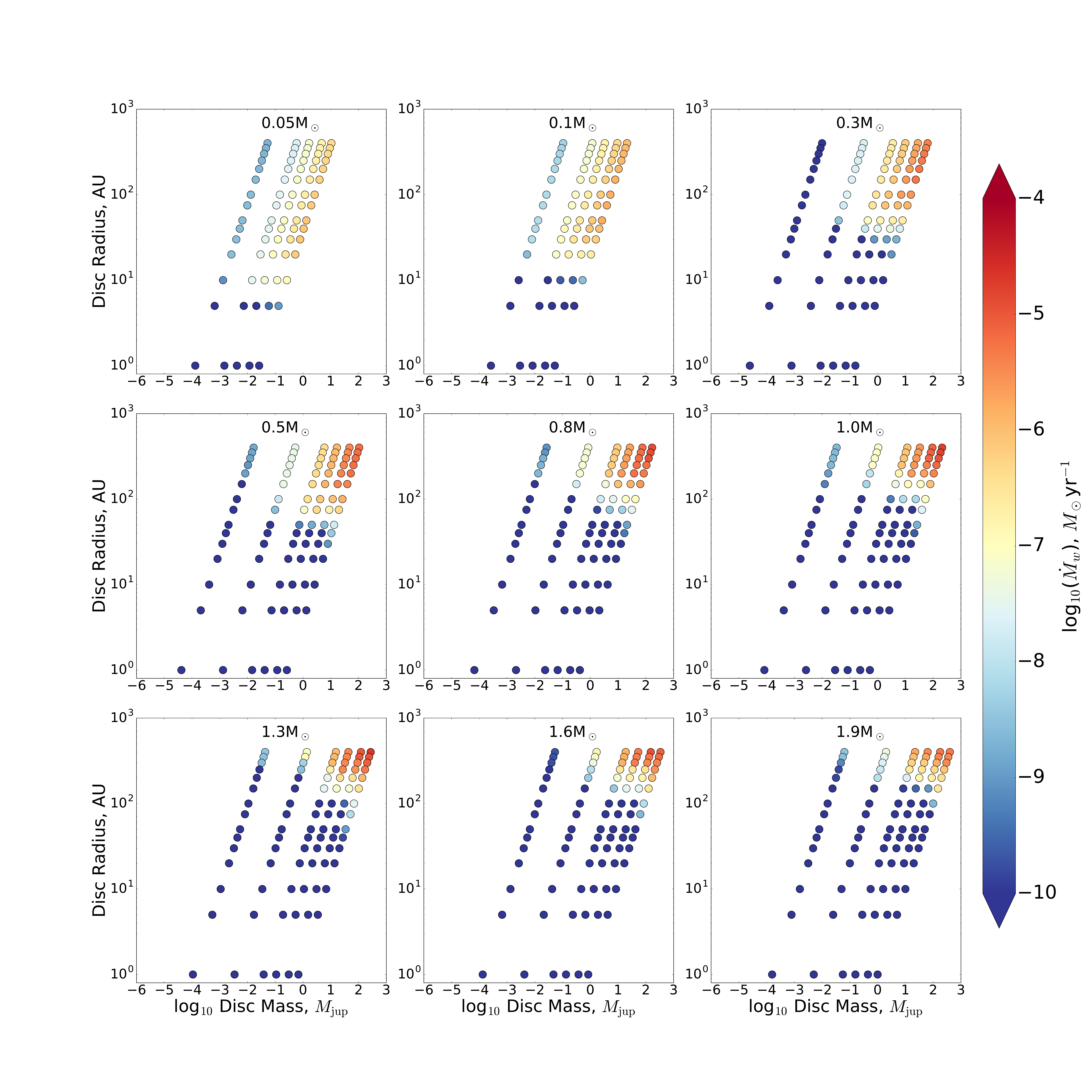}
	\caption{The mass loss rate for the $10^2$\,G$_0$ models over our entire parameter space, plotted as a function of disc mass and outer radius. The stellar mass in each case is denoted above each panel. Note that the grid has a floor value on the mass loss rate of $10^{-10}$\,M$_\odot$\,yr$^{-1}$}
	\label{fig:all100G0}
\end{figure*}

Here we summarise the \textsc{fried} mass loss rate grids in Figures \ref{fig:all10G0}--\ref{fig:all10000G0}. Each Figure is for a distinct value of the UV field, but otherwise encapsulates the mass loss rate (via the colour scale) as a function of stellar mass, disc mass and disc radius. We reiterate that the floor value in our grid is $10^{-10}$\,M$_\odot$\,yr$^{-1}$.

\begin{figure*}
	\centering	
	\includegraphics[width=19cm]{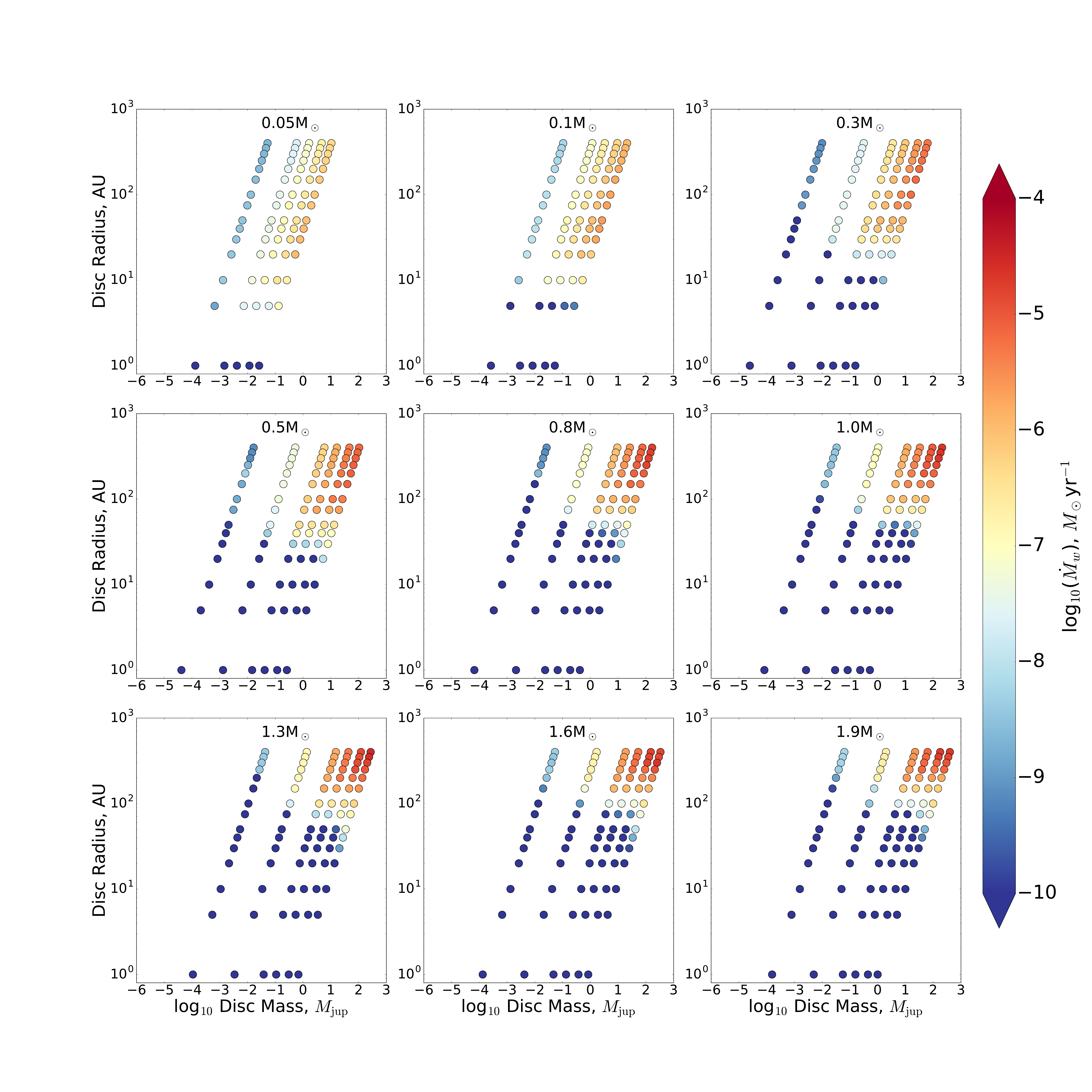}
	\caption{The mass loss rate for the $10^3$\,G$_0$ models over our entire parameter space, plotted as a function of disc mass and outer radius. The stellar mass in each case is denoted above each panel. Note that the grid has a floor value on the mass loss rate of $10^{-10}$\,M$_\odot$\,yr$^{-1}$}
	\label{fig:all1000G0}
\end{figure*}

\begin{figure*}
	\centering	
	\includegraphics[width=19cm]{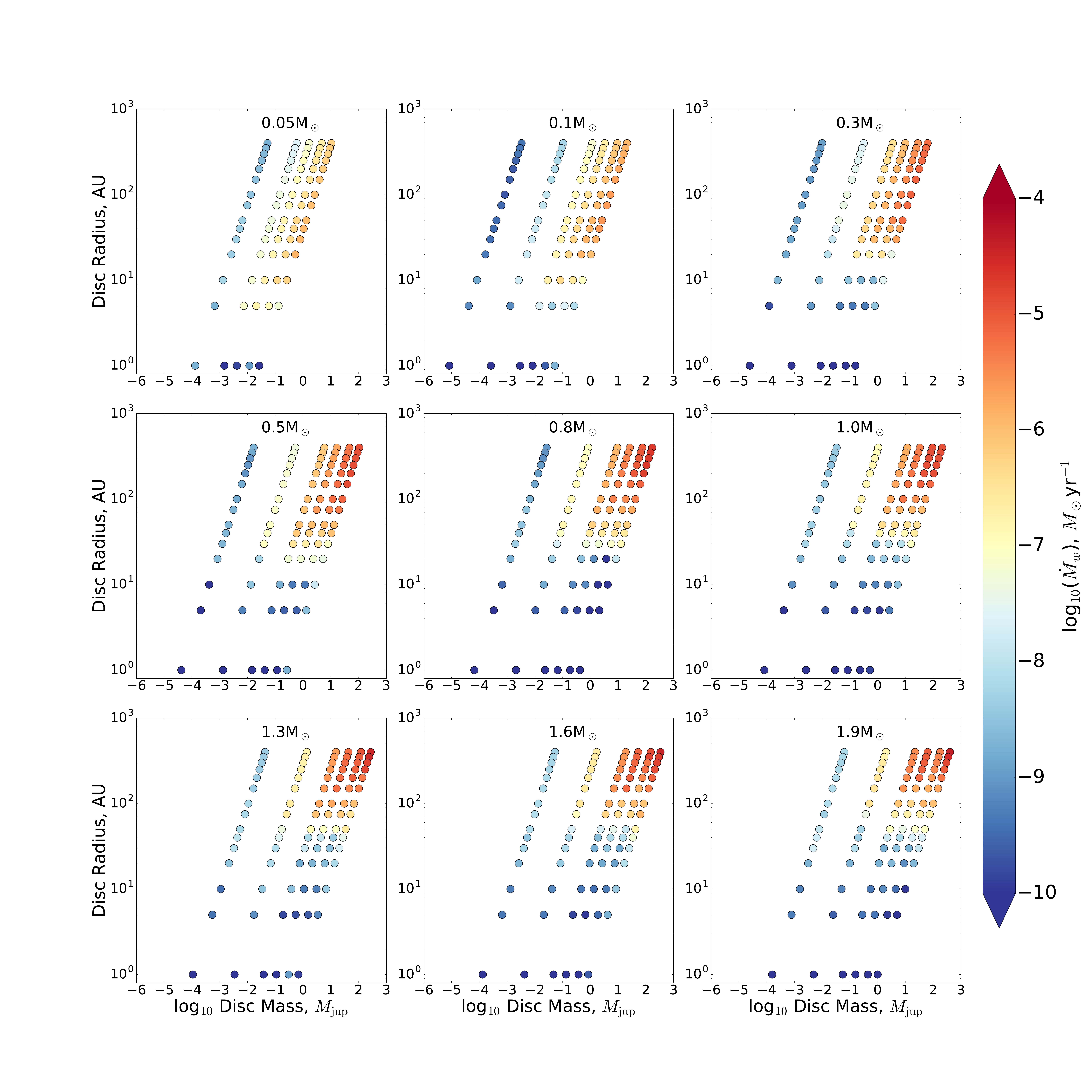}
	\caption{The mass loss rate for the $5\times10^3$\,G$_0$ models over our entire parameter space, plotted as a function of disc mass and outer radius. The stellar mass in each case is denoted above each panel. Note that the grid has a floor value on the mass loss rate of $10^{-10}$\,M$_\odot$\,yr$^{-1}$}
	\label{fig:all1000G0}
\end{figure*}

\begin{figure*}
	\centering	
	\includegraphics[width=19cm]{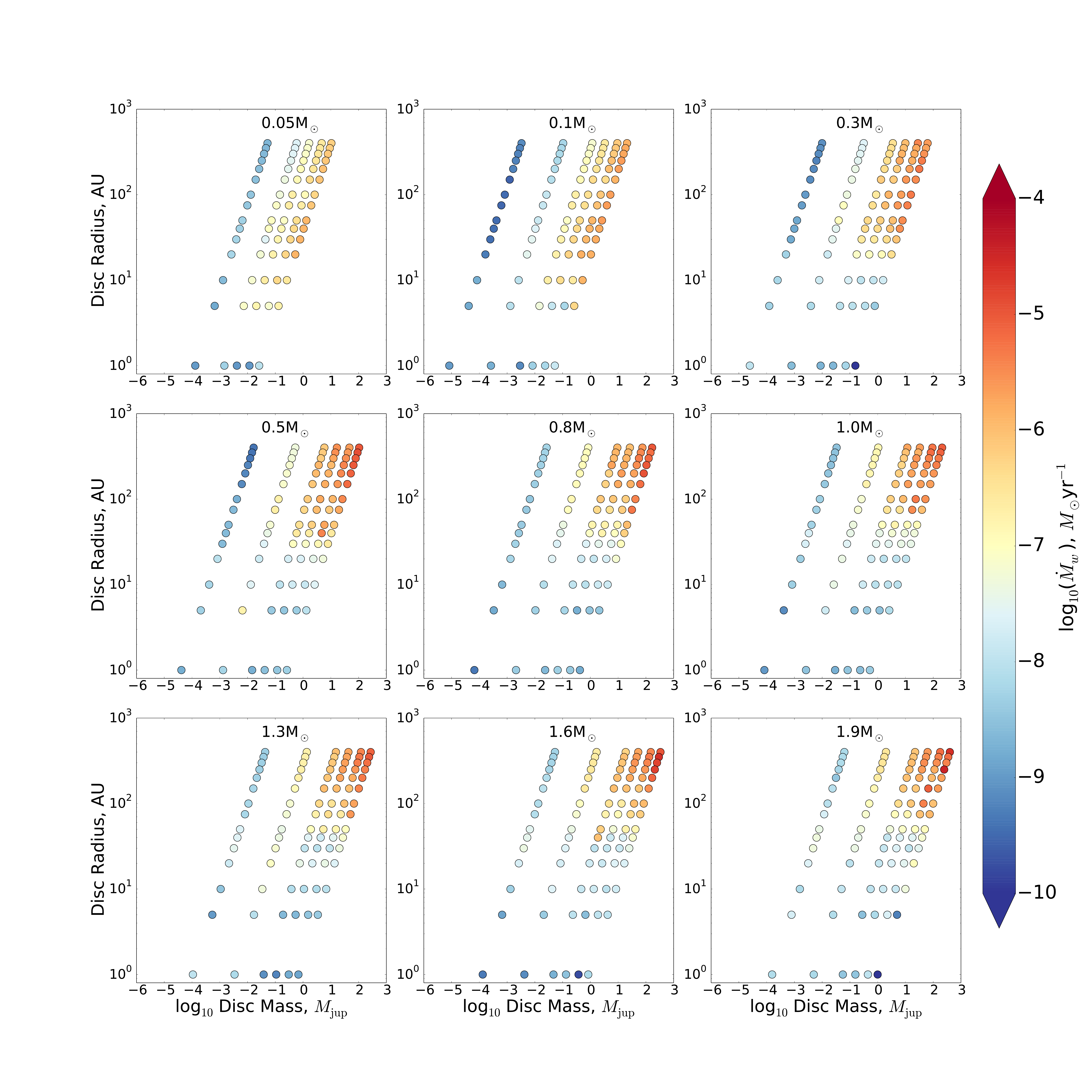}
	\caption{The mass loss rate for the $10^4$\,G$_0$ models over our entire parameter space, plotted as a function of disc mass and outer radius. The stellar mass in each case is denoted above each panel. Note that the grid has a floor value on the mass loss rate of $10^{-10}$\,M$_\odot$\,yr$^{-1}$}
	\label{fig:all10000G0}
\end{figure*}

\label{lastpage}

\end{document}